\documentclass[graphicx,preprint,amsmath,amssymb]{revtex4-1}

\usepackage[utf8]{inputenc}  
\usepackage{hyperref}
\usepackage{amsmath}
\usepackage{bm}
\usepackage{amsfonts}
\usepackage{amssymb}
\usepackage{graphicx}
\usepackage{epstopdf}
\usepackage{soul}
\usepackage{xcolor}
\newcommand{\br}{{\bf r}}

\newcommand{\bu}{{\bf u}}
\newcommand{\bq}{{\bf q}}

\newcommand{\Pt}{\widetilde{P}}

\begin{document}

\title{Noise-induced aggregation of swimmers in the Kolmogorov flow} 
\author{Simon A. Berman$^{1}$}
\email{simon.ab.berman@gmail.com}
\author{Kyle S. Ferguson\,$^{2}$}
\email{Present address: Medical Physics Graduate Program, Duke University, Durham, North Carolina 27701, USA}
\author{Nathaniel Bizzak\,$^{2}$, Thomas H. Solomon$^2$}
\author{Kevin A. Mitchell$^{1}$}
\email{kmitchell@ucmerced.edu}
\affiliation{$^{1}$Department of Physics, University of California, Merced, California 95344, USA}
\affiliation{$^{2}$Department of Physics and Astronomy, Bucknell University, Lewisburg, Pennsylvania 17837, USA}
%

\begin{abstract}
    We investigate a model for the dynamics of ellipsoidal microswimmers in an externally imposed, laminar Kolmogorov flow.  Through a phase-space analysis of the dynamics without noise, we find that swimmers favor either cross-stream or rotational drift, depending on their swimming speed and aspect ratio.  When including noise, i.e. rotational diffusion, we find that swimmers are driven into certain parts of phase space, leading to a nonuniform steady-state distribution. This distribution exhibits a transition from swimmer aggregation in low-shear regions of the flow to aggregation in high-shear regions as the swimmer’s speed, aspect ratio, and rotational diffusivity are varied.  To explain the nonuniform phase-space distribution of swimmers, we apply a weak-noise averaging principle that produces a reduced description of the stochastic swimmer dynamics. Using this technique, we find that certain swimmer trajectories are more favorable than others in the presence of weak rotational diffusion.
By combining this information with the phase-space speed of swimmers along each trajectory, we predict the regions of phase space where swimmers tend to accumulate. The results of the averaging technique are in good agreement with direct calculations of the steady-state distributions of swimmers.
In particular, our analysis explains the transition from low-shear to high-shear aggregation.
\end{abstract}
\maketitle

\section{Introduction}
The interaction between self-propelled particles and fluid flows is central to many active matter systems, including swimming bacteria \cite{Rusconi2014}, Janus particles \cite{Ebbens2010}, and microtubule-based active nematics \cite{Sanchez2012}.
A key issue for these systems is understanding how the combined effects of fluid advection and self-propulsion determine the macroscopic properties of the active suspension.
This is a nontrivial question even for dilute suspensions of active particles subjected to externally imposed, time-independent fluid flows.
In this case, connecting the behavior of individual particles to the macroscopic properties of the fluid-particle suspension may be difficult because active particles can exhibit chaotic dynamics in such flows, e.g. vortex flows \cite{Torney2007,Khurana2011,Berman2020,Ariel2020,Berman2021a}.

For sufficiently simple flows---such as time-independent planar shear flows---swimmers do not exhibit chaos, but the connection between the swimmer trajectories and the macroscopic properties of the suspension has remained elusive.
Experiments on swimming microorganisms in microfluidic channel flows reveal that the swimmer concentration profile is nonuniform across the channel and dependent on the type of microorganism and the ratio of the swimming speed relative to the flow speed \cite{Rusconi2014,Barry2015}.
Intriguingly, for certain species, the shape of the concentration profile exhibits a transition from being peaked on either side of the channel center to being peaked at the center of the channel as the flow speed increases \cite{Barry2015}.
Because the magnitude of the fluid shear is peaked at the edges of the channel and goes to zero at the center of the channel, this is characterized as a transition from low-shear depletion to high-shear depletion. 
In Ref.~\cite{Rusconi2014}, it is shown that in order to capture this key feature of the experimental data, the model for swimmer motion must take into account fluctuations in the swimming direction caused by rotational diffusion or run-and-tumble swimming.
In particular, a Fokker-Planck model \cite{Vennamneni2020} predicts a transition from low-shear to high-shear depletion in channel flows as a function of swimmer shape and rotational diffusivity, consistent with the experiments \cite{Barry2015}.

The Fokker-Planck model provides an accurate quantitative description of the swimmer density in the channel flow, but it does not provide a clear mechanism for the transition from low-shear to high-shear depletion.
In particular, it obscures the link between the concentration profile and the swimmer trajectories one actually observes in the presence of both fluid flows and rotational noise.
It also leaves open the question of how nonuniform distributions in the swimmer's phase space can arise in the first place, given that typical models of swimmer motion in planar shear flows are conservative dynamical systems perturbed by noise \cite{Zottl2012,Zottl2013,Santamaria2014}.
Because conservative systems cannot possess attractors, which typically account for the high-density regions of phase space in noisy dynamical systems, the density variations in the swimmer phase space remain unexplained.

In this paper, we study a model of an elongated swimmer in a planar Kolmogorov flow and elucidate the connection between the swimmer trajectories with noise and the swimmer density in phase space.
We choose to study the planar Kolmogorov flow---that is, a spatially periodic, alternating shear flow---rather than the channel flow, because this allows us to ignore the effect of boundary conditions on the swimmer dynamics, which can be quite complex and system dependent \cite{Chen2020}.
We calculate the steady-state distributions of swimmers in the flow numerically, which exhibit nonuniform concentration profiles in the cross-stream direction that are similar to those observed in channel flows.
In particular, we map out the transition from low- to high-shear depletion as a function of swimmer speed, shape, and rotational diffusivity.
Then, taking advantage of a conserved quantity possessed by our model, we derive a reduced drift-diffusion model for the swimmer dynamics in the limit of weak noise.
The reduced model allows us to calculate the likelihood of observing particular swimmer trajectories in the presence of noise.
Combining this with the slow-down of trajectories in phase-space, we are able to explain the nonuniformity of the swimmer steady-state distributions.
The predictions of the reduced model are quantitatively accurate in the small-diffusion limit and explain the transition from low- to high-shear depletion in terms of the relative weighting of different swimmer trajectories.

This paper is organized as follows.
In Sec.~\ref{sec:det}, we describe the phase-space structure of the swimmer in the Kolmogorov flow without noise.
In Sec.~\ref{sec:noise}, we add rotational diffusion to our model and calculate the probability density and depletion of swimmers in the Kolmogorov flow as a function of swimmer speed, shape, and rotational diffusivity.
In Sec.~\ref{sec:avg}, we introduce the reduced model for swimmer dynamics with weak diffusion, and we use the model to explain our observations from the previous section.
Concluding remarks are in Sec.~\ref{sec:concl}.
An appendix contains details of our numerical method for solving the swimmer Fokker-Planck equation.

\section{Deterministic dynamics of a swimmer in the Kolmogorov flow}\label{sec:det}
\begin{figure}
\centering
\includegraphics[width=0.6\textwidth]{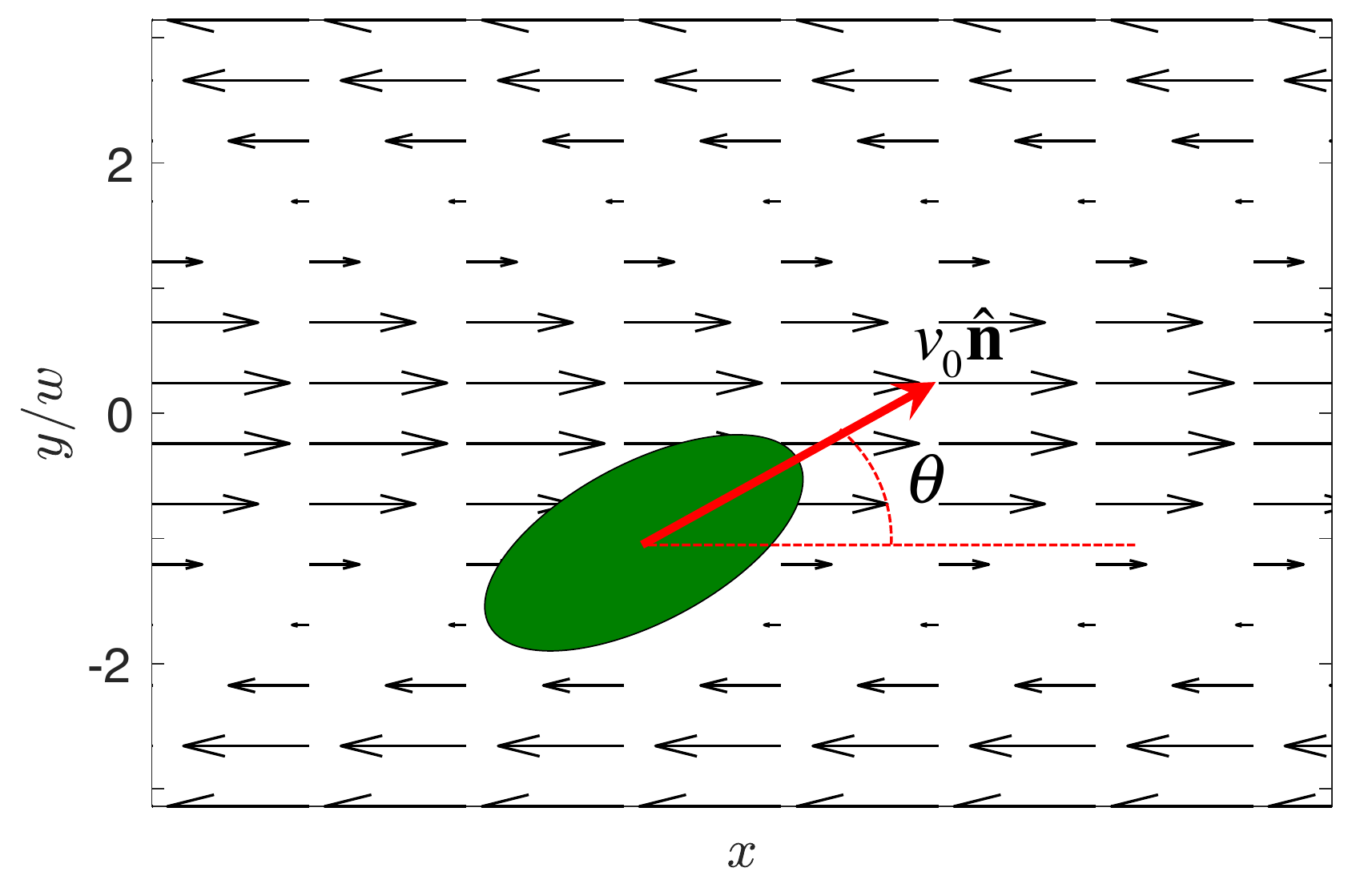}
\caption{Schematic of an ellipsoidal swimmer with shape parameter $0 < \alpha < 1$ in the Kolmogorov flow (not to scale).}\label{fig:flow}
\end{figure}
The 2D, laminar Kolmogorov flow is given by
\begin{equation}\label{eq:kolmogorov}
\bu = \left(U\cos\left( \frac{y}{w} \right),0\right).
\end{equation}
Here the constant $U$ is the maximum fluid speed at $y=0$ and $y = \pm \pi w$, where $2\pi w$ is the spatial period of the flow.
We refer to $y=0$ and $y = \pm \pi w$ as the centerlines of the flow, in analogy with the Poiseuille flow, which has maximum speed on the centerline of the channel.
In the Kolmogorov flow, the equations of motion of an ellipsoidal swimmer with position $\br = (x,y)$ and swimming direction $\theta$ relative to the $x$ axis, as shown in Fig.~\ref{fig:flow}, are \cite{Torney2007,Berman2021a}
\begin{subequations}
\begin{align}
\dot{x} & = u_x + V \cos\theta =  U\cos\left( \frac{y}{w} \right) + V \cos \theta, \\
\dot{y} & = u_y + V \sin \theta =  V \sin \theta, \\  \label{eq:thdotfull}
\dot{\theta} & = \frac{1}{2}\left(u_{y,x} - u_{x,y} \right) + \alpha \left[ \frac{1}{2} \left(u_{x,y} + u_{y,x} \right) \cos 2\theta - u_{x,x} \sin 2\theta \right]   = \frac{U}{2w}\sin\left(\frac{y}{w}\right)\left[ 1 - \alpha \cos 2\theta\right],
\end{align}
\end{subequations}
where the first equality in Eq.~\eqref{eq:thdotfull} is Jeffery's equation for an ellipsoidal particle in an incompressible flow \cite{Jeffery1922}.
We have used the notation $(\cdot)_{,x} \equiv \partial (\cdot) / \partial x$.
The parameter $V$ is the swimming speed, and $\alpha = (\gamma^2 - 1)/(\gamma^2 + 1)$ is the swimmer shape parameter, where $\gamma$ is the ellipsoid aspect ratio.
The parameter $\alpha$ takes values between $-1$ and $1$, with $\alpha = 1$ corresponding to a rod swimming parallel to its major axis, $\alpha = 0$ corresponding to a circular swimmer, and $\alpha = -1$ corresponding to a rod swimming perpendicular to its major axis. 
A schematic of the model is shown in Fig.~\ref{fig:flow}.

In this paper, we focus on the dynamics in the $(y,\theta)$ plane, since the system is translationally invariant in the $x$ direction.
Furthermore, we work with the non-dimensionalized version of Eq.~\eqref{eq:kolmogorov}, which we obtain by normalizing by the velocity scale $U$ and the length scale $w$.
This leads to
\begin{subequations}\label{eq:kolmogorov_nondim}
\begin{align}\label{eq:ydot}
\dot{y} & = v_0 \sin \theta, \\ \label{eq:thdot}
\dot{\theta} & = \frac{\sin y}{2}\left[ 1 - \alpha \cos 2\theta\right],
\end{align}
\end{subequations}
with the rescaled swimming speed $v_0 = V/U$.
Because Eq.~\eqref{eq:kolmogorov_nondim} is periodic in both $y$ and $\theta$, we interpret the phase space  topologically as a two torus, i.e. a donut.
Remarkably, the system \eqref{eq:kolmogorov_nondim} is conservative, as is the dynamics of ellipsoidal swimmers in 2D and 3D Poiseuille flows \cite{Zottl2012,Zottl2013}, spherical gyrotactic swimmers in Kolmogorov flows \cite{Santamaria2014}, and spherical swimmers in 2D vortex flows \cite{Arguedas-Leiva2020}.
The conserved quantity may be derived in the standard way.
Dividing Eq.~\eqref{eq:ydot} by Eq. \eqref{eq:thdot}, we obtain
\begin{equation}\label{eq:dydtheta}
\frac{{\rm d} y}{{\rm d} \theta} = \frac{2 v_0 \sin \theta}{\sin y \left(1 - \alpha \cos 2\theta \right)}.
\end{equation}
Separating variables in Eq.~\eqref{eq:dydtheta}, we obtain
\begin{equation}\label{eq:sepvar}
\sin y \,{\rm d} y = \frac{2 v_0 \sin \theta}{1 - \alpha \cos 2\theta} {\rm d} \theta.
\end{equation}
Equation \eqref{eq:sepvar} may be integrated, which introduces an integration constant $\Psi$.
This implies that $\Psi(y,\theta)$ is a constant of motion, and it is given by
\begin{equation}\label{eq:Psi}
\Psi(y,\theta) = \begin{cases}
\cos y - 2v_0\frac{\tanh^{-1}\left(\sqrt{\frac{2\alpha}{1+\alpha}} \cos\theta \right)}{\sqrt{2\alpha(1+\alpha)}}, & {\rm  for}\;  0 < \alpha \leq 1, \\
\cos y - 2v_0 \cos \theta, & {\rm  for }\; \alpha = 0, \\
\cos y - 2v_0 \frac{\tan^{-1}\left(\sqrt{\frac{2|\alpha|}{1+\alpha}} \cos\theta \right)}{\sqrt{2|{\alpha}|(1+\alpha)}}, & {\rm  for }\;  -1 < \alpha <  0.
\end{cases}
\end{equation}
\begin{figure}
\centering
\includegraphics[width=\textwidth]{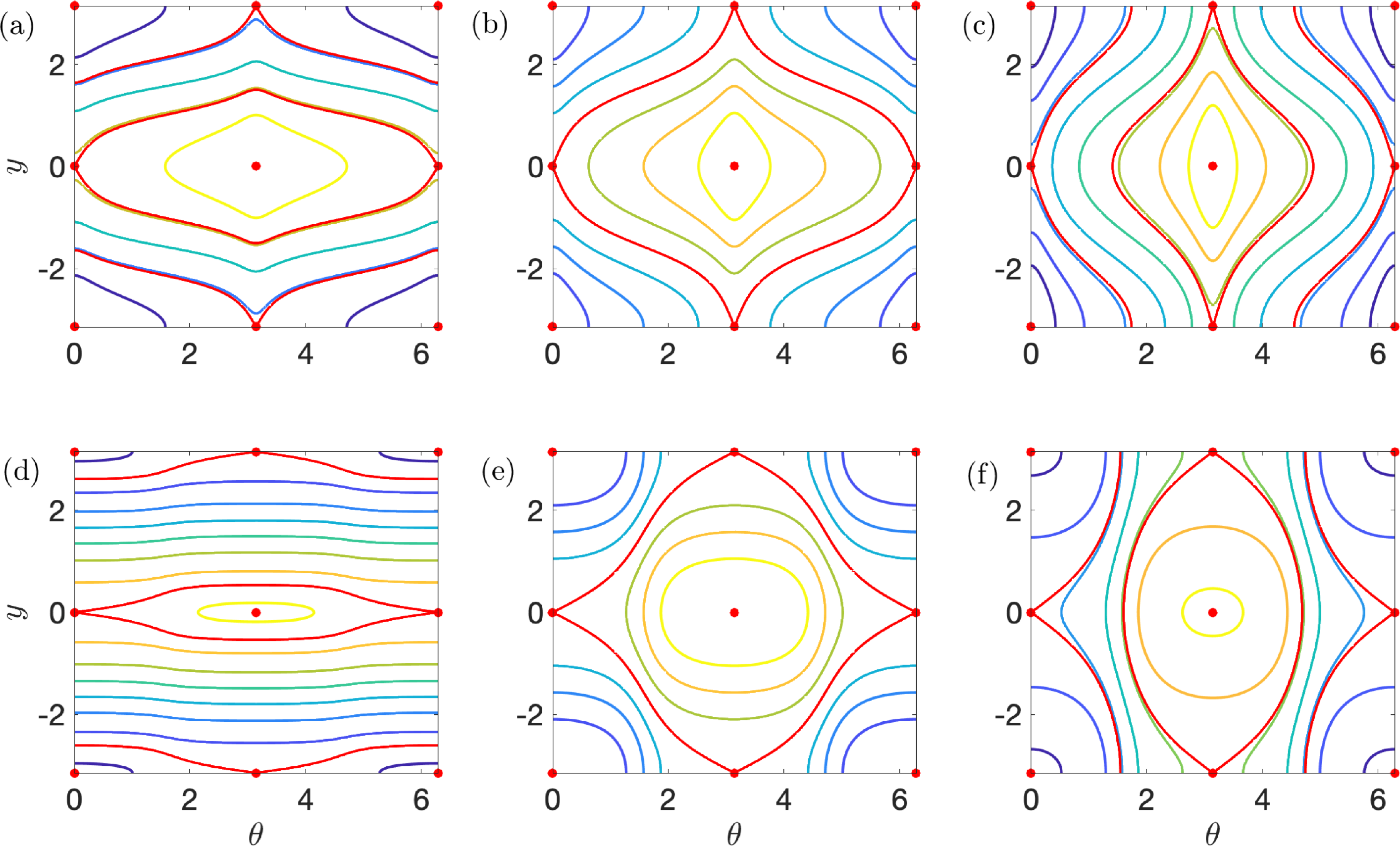}
\caption{Phase portraits of Eq.~\eqref{eq:kolmogorov_nondim} for (a--c) $\alpha = 0.9$ and (d--f) $\alpha = -0.7$. (a) $v_0 = 0.2$. (b) $v_0 = v^*(\alpha) = 0.430$. (c) $v_0 = 0.8$. (d) $v_0 = 0.02$.  (e) $v_0 = v^*(\alpha) = 0.285$. (f) $v_0 =  0.6$. Red dots are the fixed points of  Eq.~\eqref{eq:kolmogorov_nondim}. Red curves are the separatrices emanating out of the saddle fixed points at $(y,\theta) = (0,0)$ and $(y,\theta) = (\pm \pi,\pi)$. The color of each remaining curve corresponds to a distinct value of $\Psi$.}\label{fig:det_phase_portrait}
\end{figure}

As a result, each trajectory of Eq.~\eqref{eq:kolmogorov_nondim} lies on a constant-$\Psi$ contour and the phase space is foliated by periodic trajectories almost everywhere.
Illustrative phase portraits are shown in Figs.~\ref{fig:det_phase_portrait}a--c for $\alpha = 0.9$ and Figs.~\ref{fig:det_phase_portrait}d--f for $\alpha = -0.7$. 
The phase portraits are organized around the four fixed points of Eq.~\eqref{eq:kolmogorov_nondim}.
For all $v_0$ and all $\alpha < 1$, the fixed points are $(y_1,\theta_1) = (0,0)$, $(y_2,\theta_2) = (0,\pi)$, $(y_3,\theta_3) = (\pi,\pi)$ and $(y_4,\theta_4) = (\pi,0)$.
The fixed points $(y_1,\theta_1)$ and $(y_3,\theta_3)$ are saddles, while $(y_2,\theta_2)$ and $(y_4,\theta_4)$ are centers, and the two fixed points within each pair map into each other by the shift-flip symmetry
\begin{equation}\label{eq:symmshift}
(y,\theta) \mapsto (y+\pi,\pi - \theta)
\end{equation}
of Eq.~\eqref{eq:kolmogorov_nondim}.
The centers are surrounded by periodic orbits, as shown in Fig.~\ref{fig:det_phase_portrait}.
This can either be viewed as a consequence of the conservative nature of the system, or a consequence of the time-reversal symmetries
\begin{subequations}\label{eq:tsymm}
\begin{align}
(y,\theta,t) & \mapsto (y,-\theta,-t), \\
(y,\theta,t) & \mapsto (-y,\theta,-t).
\end{align} 
\end{subequations}
Indeed, all of the fixed points are invariant under both of the symmetries \eqref{eq:tsymm}, in particular the centers, which guarantees that they are surrounded by periodic orbits \cite{Strogatz}.

The time-reversal symmetries \eqref{eq:tsymm} also force the invariant manifolds of the saddles to be homoclinic orbits.
These homoclinic orbits (red curves in Fig.~\ref{fig:det_phase_portrait}) act as separatrices between two topologically distinct types of periodic trajectories (Figs.~\ref{fig:det_phase_portrait}a, \ref{fig:det_phase_portrait}c, \ref{fig:det_phase_portrait}d, and \ref{fig:det_phase_portrait}f).
Either (i) a trajectory is enclosed by the separatrices of a saddle, and hence wraps around the center inside the separatrices, or (ii) the trajectory is in between the separatrices of the two saddles, and hence does not wrap around a center.
In other words, the homoclinc orbits separate orbits which are (i) contractible on the torus from those which are (ii) non-contractible.
We refer to the regions enclosed by a saddle's separatrices as the islands, and the remaining regions as jets.

For sufficiently small $v_0$ (Figs.~\ref{fig:det_phase_portrait}a and \ref{fig:det_phase_portrait}d), the spatially varying shear suppresses sustained cross-stream drift and promotes rotational drift.
The islands contain swimmer trajectories that oscillate around the centerlines, while the swimmer's orientation oscillates back and forth, never completing a $2\pi$ rotation.
Meanwhile, the jets contain swimmer trajectories that oscillate in the $y$ direction, never crossing the centerlines, while the swimmer orientation continuously drifts either clockwise or counterclockwise.
On the other hand, for sufficiently large $v_0$, (Figs.~\ref{fig:det_phase_portrait}c and \ref{fig:det_phase_portrait}f), a swimmer's self-propulsion can prevail and give rise to sustained motion in the $y$-direction.
While the islands are qualitatively similar to the small $v_0$ case, the jets now either transport swimmers in the $+y$ or $-y$ direction, depending on $\theta$.
In these jets, a swimmer's orientation never completes a full rotation, but oscillates around $\theta = \pi/2$ or $\theta = 3\pi/2$, respectively.

\begin{figure}
\centering
\includegraphics[width=0.5\textwidth]{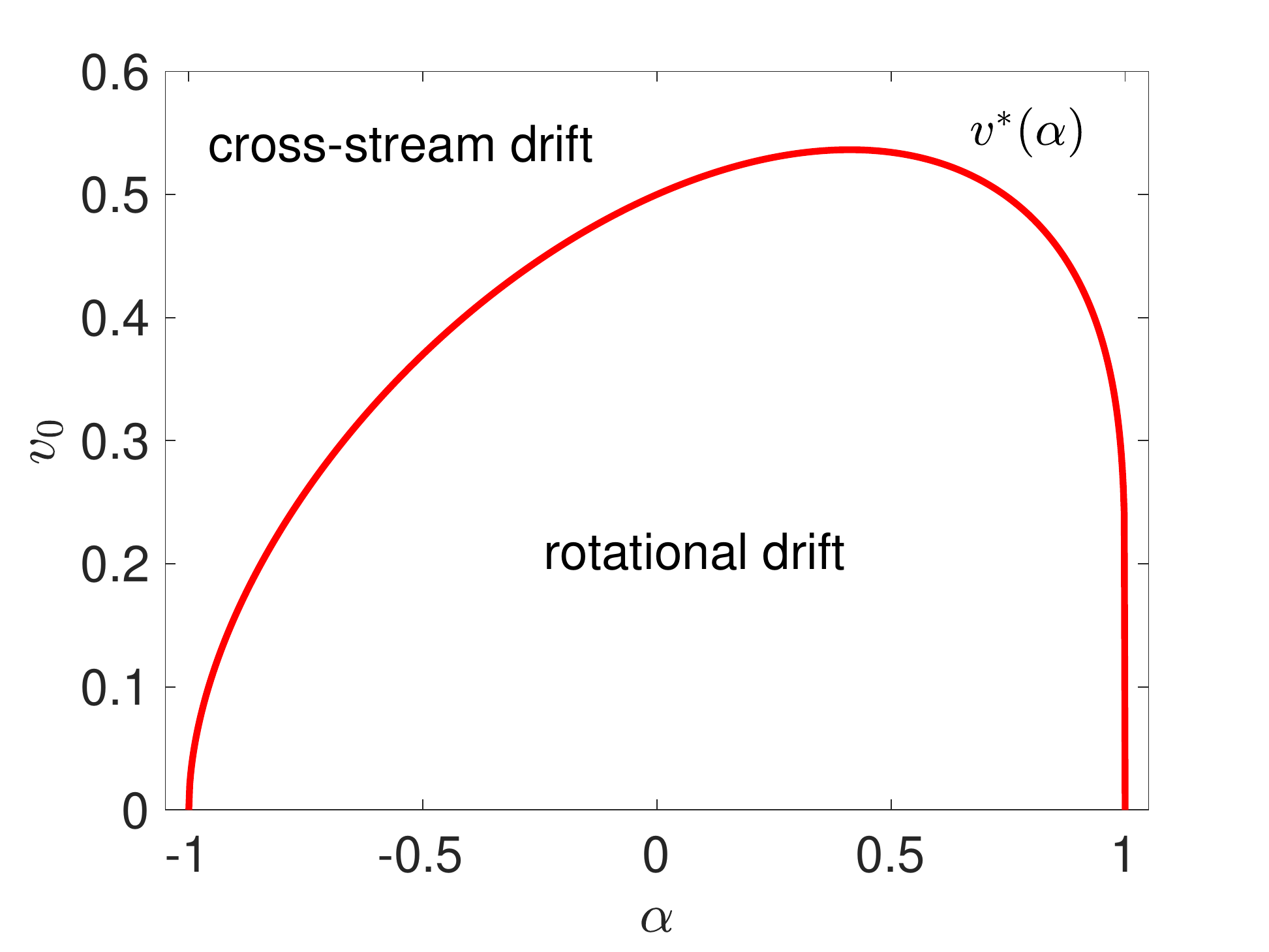}
\caption{Swimmer phase space structure as a function of the swimming speed $v_0$ and shape parameter $\alpha$. The red curve is $v^*(\alpha)$ (Eq.~\eqref{eq:vBif}), the set of parameters where the saddles are heteroclinically connected and phase space bifurcates from containing rotational drift trajectories to cross-stream drift trajectories.}\label{fig:vBif}
\end{figure}
The transition from rotational drift to cross-stream drift in the jets occurs at a shape-dependent swimming speed $v^*(\alpha)$, at which the system \eqref{eq:kolmogorov_nondim} exhibits a global bifurcation.
The bifurcation takes place when the separatrices of each of the saddle points coincide exactly, as shown in Figs.~\ref{fig:det_phase_portrait}b and \ref{fig:det_phase_portrait}e. 
This occurs when the level sets containing the saddles $(0,0)$ and $(\pi,\pi)$ are identical, meaning $\Psi(0,0) = \Psi(\pi,\pi)$.
Using Eq.~\eqref{eq:Psi}, this relation leads to $v^*(\alpha)$, given by
\begin{equation}\label{eq:vBif}
v^*(\alpha) = \begin{cases}
        \frac{\sqrt{2\alpha(1+\alpha)}}{2\tanh^{-1}\left(\sqrt{\frac{2\alpha}{1+\alpha}}\right)}, & \text{for } 0 < \alpha < 1, \\
        \frac{1}{2}, & \text{for } \alpha = 0,\\
        \frac{\sqrt{2|\alpha|(1+\alpha)}}{2\tan^{-1}\left(\sqrt{\frac{2|\alpha|}{1+\alpha}}\right)}, & \text{for } -1 \leq \alpha < 0.
        \end{cases}
\end{equation}
The bifurcation curve is plotted in Fig.~\ref{fig:vBif}.
This transition from rotational to cross-stream drift is an interesting aspect of the swimmer dynamics in the Kolmogorov flow, which has no direct counterpart in channel flows.
A similar phenomenon has been observed numerically for gyrotactic swimmers in the Kolmogorov flow \cite{Santamaria2014}.

\section{Stochastic dynamics and noise-driven aggregation}\label{sec:noise}
In both laboratory experiments and nature, real swimmers are subject to fluctuations in their motion which cause their trajectories to deviate from Eq.~\eqref{eq:kolmogorov_nondim}.
We consider the situation where the swimming direction Eq.~\eqref{eq:thdot} is perturbed by white noise, while the translational motion continues to follow Eq.~\eqref{eq:ydot}.
This description is a simple model for biological microswimmers, like swimming bacteria.
For such swimmers, translational fluctuations are negligible compared to their self-propulsion, but rotational fluctuations are significant due to the fluctuating active forces that propel the swimmer \cite{Thiffeault2021,Hyon2012} and potentially run-and-tumble behavior as well \cite{Rusconi2014}.
Equation \eqref{eq:kolmogorov_nondim} is replaced by the stochastic differential equation
\begin{subequations}\label{eq:kolmogoroVde}
\begin{align}\label{eq:ydot_sde}
{\rm d}y  & = v_0 \sin \theta {\rm d} t, \\ \label{eq:thdot_sde}
{\rm d} \theta & = \frac{\sin y}{2}\left[ 1 - \alpha \cos 2\theta\right] {\rm d} t + \sigma {\rm d} W,
\end{align}
\end{subequations}
where $\sigma$ is the dimensionless strength of rotational diffusion.
In Eq.~\eqref{eq:thdot_sde}, ${\rm d}W$ is the infinitesimal increment of a Wiener process, meaning for every instant of time $t$, ${\rm d}W$ is a zero-mean normally-distributed random variable with variance ${\rm d}t$.
We denote a particular realization of the Wiener process as $W(t) = \int_0^t {\rm d} W$.
The noise strength $\sigma$ can be expressed in terms of the rotational diffusivity $D_R$ as
\begin{equation}
\sigma = \sqrt{\frac{2 D_R w}{U}}.
\end{equation}

\subsection{Steady-state probability density}
\begin{figure}
\centering
\includegraphics[width=\textwidth]{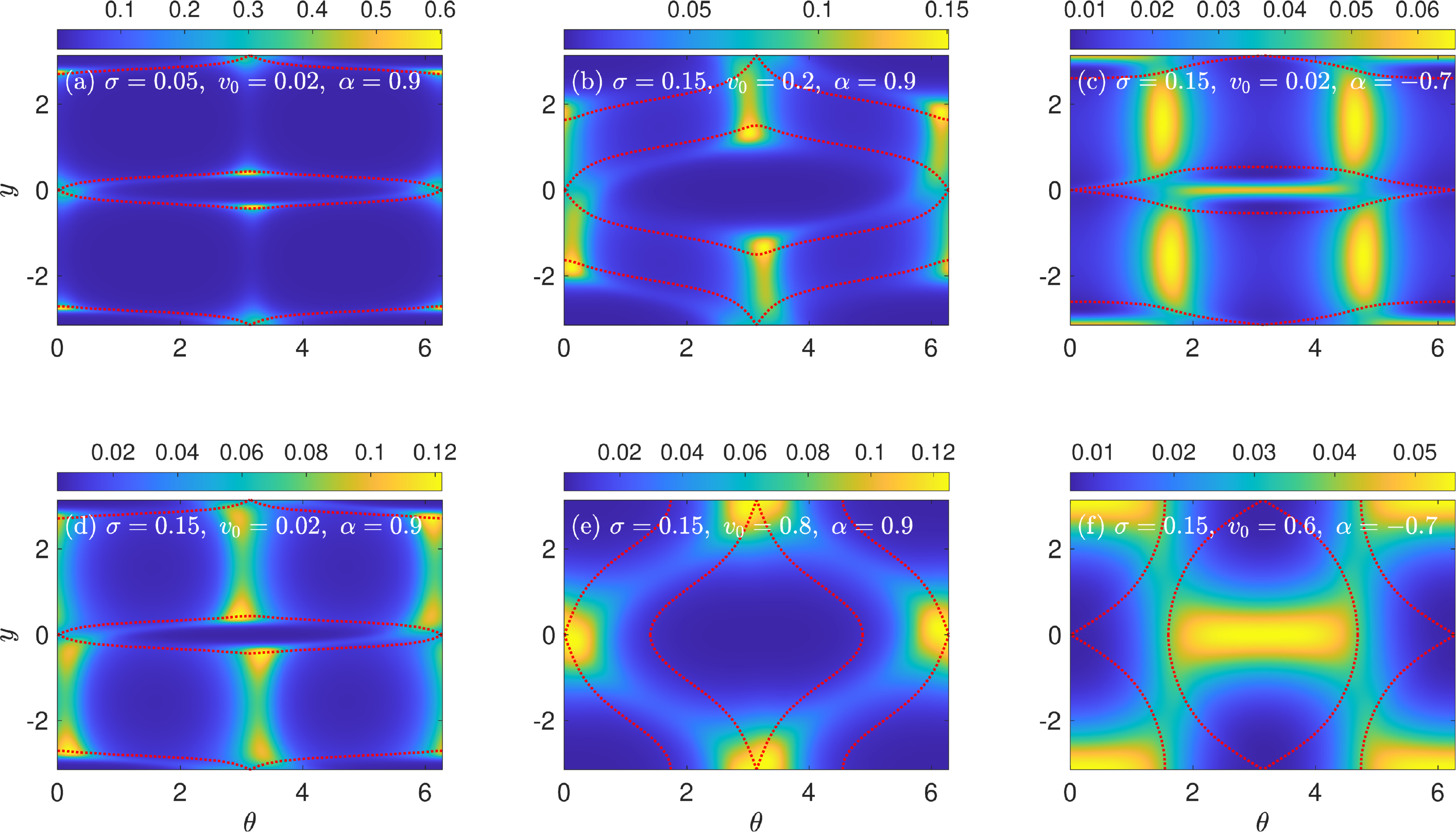}
\caption{Steady-state probability densities $P(y,\theta)$ for various combinations of $\sigma$, $v_0$, and $\alpha$ indicated on each panel. Red dotted curves are the separatrices emanating from the saddles $(y,\theta) = (0,0)$ and $(y,\theta) = (\pm \pi,\pi)$. Panels (a-d) are for parameters before the bifurcation from rotational to cross-stream drift, and panels (e) and (f) are after this bifurcation.}\label{fig:Pyth}
\end{figure}
Of particular interest is the steady-state probability density $P(y,\theta)$ of Eq.~\eqref{eq:kolmogoroVde}, which has been measured in experiments on the channel flow \cite{Rusconi2014,Barry2015}.
The density $P$ satisfies the steady-state Fokker-Planck equation associated with Eq.~\eqref{eq:kolmogoroVde} \cite{Solon2015,Berman2021b},
\begin{equation}\label{eq:fp}
- v_0\sin\theta \frac{\partial P}{\partial y} -\frac{\sin y}{2}  \frac{\partial}{\partial \theta} \left[ \left(1 - \alpha \cos 2\theta \right) P \right] + \frac{\sigma^2}{2} \frac{\partial^2 P}{\partial \theta^2} = 0.
\end{equation}
For circular swimmers,  $\alpha = 0$ and Eq.~\eqref{eq:fp} is linear in the partial derivatives of $P$.
Hence, the uniform distribution $P = {\rm const}$ is a solution, and for $\sigma \neq 0$, we suspect that this solution is unique.
Therefore, perfectly circular swimmers with rotational diffusion attain a uniform distribution in phase space.
For $\alpha \neq 0$, the periodicity of Eq.~\eqref{eq:fp} in $y$ and $\theta$ means it is straightforward to solve numerically using Fourier transforms.
This aspect also makes the Kolmogorov flow more convenient to work with than a channel flow.
We describe our numerical method in Appendix \ref{sec:app}, and our code is available in Ref.~\cite{code}.
Example solutions are plotted in Fig.~\ref{fig:Pyth}.
The plots are centered on one of the stable equilibria, $(y,\theta) = (0,\pi)$, and the separatrices enclosing each island are shown as the dotted red curves.
The density inside the central $y \in [-\pi/2,\pi/2]$ region corresponds to rightward fluid flow (in the $x$ direction), and it is related to the density in the leftward-flow region by the symmetry \eqref{eq:symmshift}.
When $\alpha \neq 0$, Fig.~\ref{fig:Pyth} shows that $P(y,\theta)$ is highly nonuniform, featuring sharp peaks in certain regions.

For $\alpha > 0$ (i.e. swimming parallel to the particle's long axis) and swimming speeds below the bifurcation [$v_0 < v^*(\alpha)$, Figs.~\ref{fig:Pyth}a, \ref{fig:Pyth}b, and \ref{fig:Pyth}d], the density is mostly concentrated near $\theta = 0$ and $\pi$, with the $y$ position of each peak lying close to the separatrices between the islands and the jets.
At these angles, the deterministic part of the angular velocity [Eq.~\eqref{eq:thdot_sde}] is minimized, meaning the swimmer spends more time near $\theta = 0$ and $\pi$.
Hence, this explains the increased density around those angles.
When the noise is small (Fig.~\ref{fig:Pyth}a), there are three peaks around each island: one at the saddle point at $\theta = 0$ (which appears as two peaks in Fig.~\ref{fig:Pyth} due to our cutting the torus at $\theta = 0$), and one on each side of the separatrix.
Meanwhile, the density in the center of the islands, around the center equilibrium points, is small and at a local minimum.
This suggests that the centers are unstable with respect to perturbations from rotational diffusion, which is not {\it a priori} obvious since centers are linearly stable.
The small-noise distribution (Fig.~\ref{fig:Pyth}a)  approximately respects the time-reversal symmetries \eqref{eq:tsymm}.
As rotational diffusion increases (Fig.~\ref{fig:Pyth}d), the peak around the saddle point splits into two peaks, and the distributions increasingly break the time-reversal symmetries.
This is evidenced by the asymmetry of Fig.~\ref{fig:Pyth}d with respect to reflections about $y = 0$ and reflections about $\theta = \pi$.
The increased noise also causes the peaks near the saddle at $(y,\theta)=(0,0)=(0,2\pi)$ to begin merging with the peaks along the middle of the separatrix surrounding the island centered on $y = \pm \pi$.
This merging becomes more pronounced for higher $v_0$, because the separatrices get closer together (Fig.~\ref{fig:Pyth}b).

For $\alpha > 0$ and swimming speeds above the bifurcation (Fig.~\ref{fig:Pyth}e), the density is peaked on both of the saddles, with most of the remaining density concentrated in the cross-stream jet region.
The occurrence of peaks on the saddles may be surprising, because these fixed points are linearly unstable and hence one might not expect density to accumulate nearby.
Again, the density inside the islands is small.
Hence, we can sum up the behavior of $\alpha > 0$ swimmers as follows.
The effect of small noise is to eject swimmers from the centers of the islands and make them aggregate near the separatrices.
Increasing the intensity of noise tends to preferentially push the swimmer density into the jet region, as seen when comparing Figs.~\ref{fig:Pyth}a and \ref{fig:Pyth}d, but the density is still strongly peaked at specific angles near $\theta = 0$ and $\pi$.
The aggregation of swimmer density near the separatrices is consistent with similar behavior seen in Monte Carlo simulations of swimmers in Poiseuille flow \cite{Rusconi2014}.
These effects are more pronounced the more elongated a swimmer is, i.e.\ the closer  $\alpha$ is to $1$.
As $\alpha \rightarrow 0$ and the swimmer shape becomes more circular, the distribution $P(y,\theta)$ tends towards the uniform distribution, so the aggregation effects gradually diminish.

On the other hand, for $\alpha < 0$ (i.e.\ swimming perpendicular to the particle's long axis), the density is peaked both at the center fixed points and away from the islands at the angles $\theta = \pi/2 $ and $3\pi/2$ (Figs.~\ref{fig:Pyth}c and \ref{fig:Pyth}f).
When $v_0$ is small and below the bifurcation, the peaks around the centers have a narrow width in the $y$ direction, as seen in Fig.~\ref{fig:Pyth}c.
Meanwhile, there are prominent peaks for  $\theta = \pi/2 $ and $3\pi/2$ in the jet region.
These angles correspond to the minimal angular speeds in Eq.~\eqref{eq:thdot_sde} for the $\alpha < 0$ case, where the swimmers spend more time.
The peaks around the centers suggest that for $\alpha < 0$, the centers are metastable with respect to perturbations from rotational diffusion.
This is a stark difference from the $\alpha > 0$ case, for which $P(y,\theta)$ is at a local minimum at the centers, and it is all the more striking because the linear stability type of the center fixed points does not depend on $\alpha$.
As $v_0$ increases, the peaks around the centers intensify while the peaks at $\theta = \pi/2 $ and $3\pi/2$ gradually fade.
For a sufficiently large $v_0$ (Fig.~\ref{fig:Pyth}f), the peaks at the centers completely dominate the probability distribution.
Though we have selected a $v_0$ above the bifurcation in Fig.~\ref{fig:Pyth}f, the peaks at the centers become dominant compared to those at $\theta = \pi/2 $ and $3\pi/2$ for values of $v_0$ well below the bifurcation.
This behavior is revealed by examining the swimmer cross-stream concentration profiles.

\subsection{Cross-stream concentration profile and depletion}
\begin{figure}
\centering
\includegraphics[width=\textwidth]{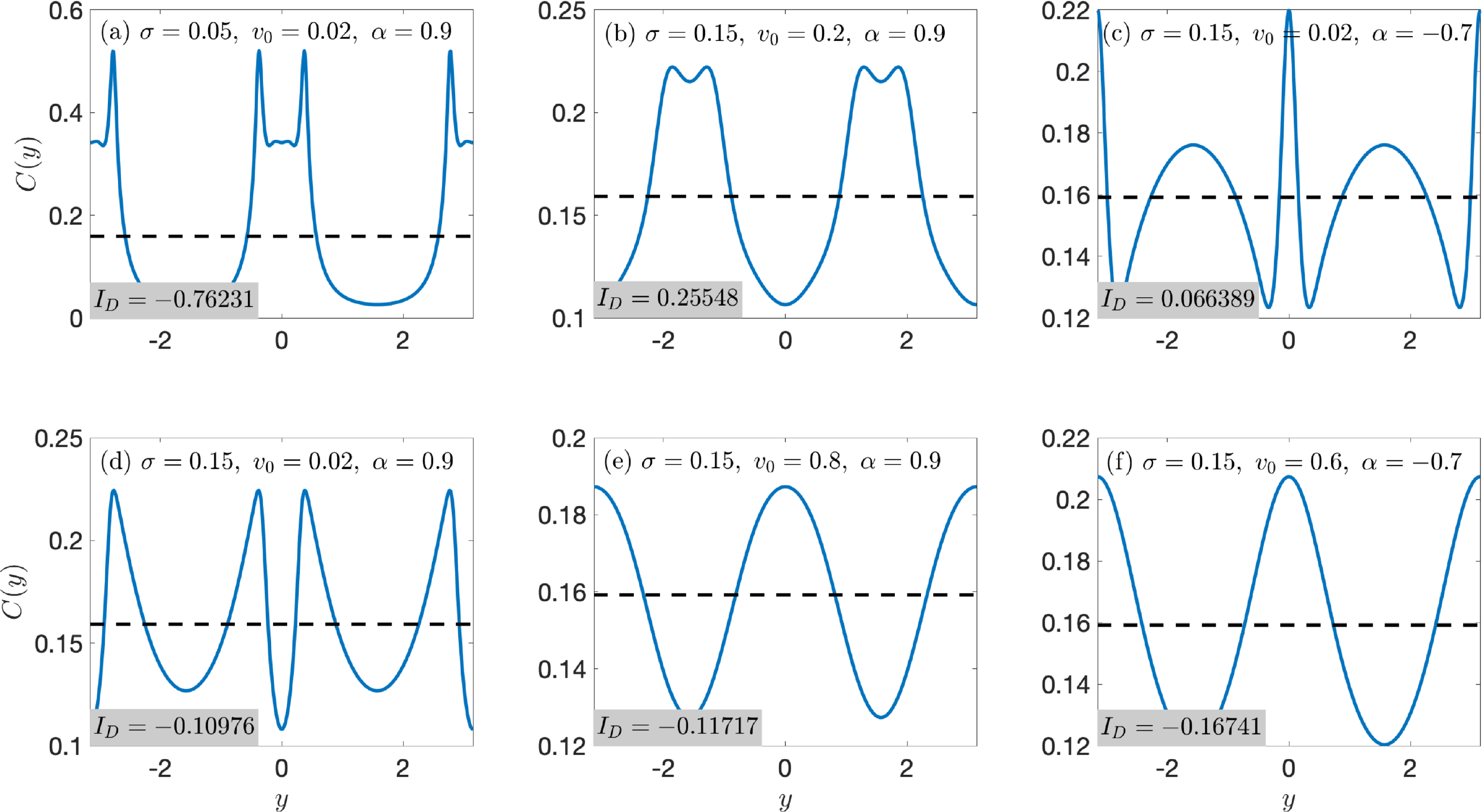}
\caption{Concentration profiles $C(y)$ for various combinations of $\sigma$, $v_0$, and $\alpha$ corresponding to those used in Fig.~\ref{fig:Pyth}. The dashed lines show the uniform concentration profile, to highlight the nonuniformity of the swimmer concentrations $C(y)$. The depletion index $I_D$ for each parameter is noted on each panel.}\label{fig:Cy}
\end{figure}
Using the calculated densities $P(y,\theta)$, we investigate the depletion of swimmers from low- or high-shear regions of the flow, which is known to exhibit a complex dependence on the swimmer parameters $(v_0,\alpha,\sigma)$.
Following prior work \cite{Rusconi2014,Barry2015,Vennamneni2020}, we define a depletion index to quantify the nonuniformity of the spatial distribution of swimmers.
The spatial distribution is given by
\begin{equation}
C(y) = \int_0^{2\pi} P(y,\theta){\rm d} \theta.
\end{equation}
The depletion index $I_D$ is defined by computing the ratio of the concentration in the low-shear regions of the flow to the uniform concentration, and subtracting this from $1$.
For Poiseuille flows, the cutoff for the low-shear region is usually taken to be the central half-width of the channel.
In the case of the Kolmogorov flow, which is like a periodically alternating Poiseuille flow, we take a similar definition as the central half-width of a region in which the flow points in one direction.
Hence, we define $I_D$ as
\begin{equation}\label{eq:Id}
I_D = 1 - 2 \left( \int_{-\pi/4}^{\pi/4} C(y) {\rm d} y + \int_{3\pi/4}^{\pi} C(y) {\rm d} y + \int_{-\pi}^{-3\pi/4} C(y) {\rm d} y \right).
\end{equation}
The first term in Eq.~\eqref{eq:Id} is the concentration in the rightward low-shear region, and the second two terms comprise the concentration in the leftward low-shear region (see Fig.~\ref{fig:flow}).
For a uniform density, $I_D = 0$; for low-shear depletion of swimmers, $I_D > 0$; for high-shear depletion, $I_D < 0$.

Example results of these calculations are shown in Fig.~\ref{fig:Cy} for the parameters used in Fig.~\ref{fig:Pyth}.
We observe a diverse set of concentration profiles that reflect the strong dependence of the density $P(y,\theta)$ on the system parameters.
For $\alpha > 0$ and small $v_0$ (Figs.~\ref{fig:Cy}a and \ref{fig:Cy}d), $C(y)$ is peaked on either side of each centerline as a result of the large peaks along the separatrices seen in Figs.~\ref{fig:Pyth}a and \ref{fig:Pyth}d.
Because the islands are very narrow in the $y$ direction when $v_0$ is small, the concentration peaks on either side of the centerline are close together.
Hence, we obtain a negative depletion index, indicating high-shear depletion.
For higher values of $v_0$, the separatrices [and the peaks of $P(y,\theta)$] move away from the centerlines (Fig.~\ref{fig:Pyth}b), leading to a positive depletion index  (Fig.~\ref{fig:Cy}b).
However, when $v_0$ is sufficiently large and past the bifurcation, the accumulation of density on the saddles  at $y = 0$ and $\pm \pi$ (Fig.~\ref{fig:Pyth}e) leads again to a negative depletion index (Fig.~\ref{fig:Cy}e).
For $\alpha < 0$ and small $v_0$ (Fig.~\ref{fig:Cy}c), the concentration profile has peaks both at the centerlines and in the high-shear regions, owing to the $P(y,\theta)$ peaks at the centers and the jet regions, respectively (Fig.~\ref{fig:Pyth}c).
Whether one observes a positive or negative $I_D$ depends on the relative prominence of these two sets of peaks in phase space.
As $v_0$ increases and the peaks at the centers dominate (Fig.~\ref{fig:Pyth}f), the swimmer concentration only features peaks at the centerlines (Fig.~\ref{fig:Cy}f), leading to a negative depletion index.

\begin{figure}
\centering
\includegraphics[width=0.6\textwidth]{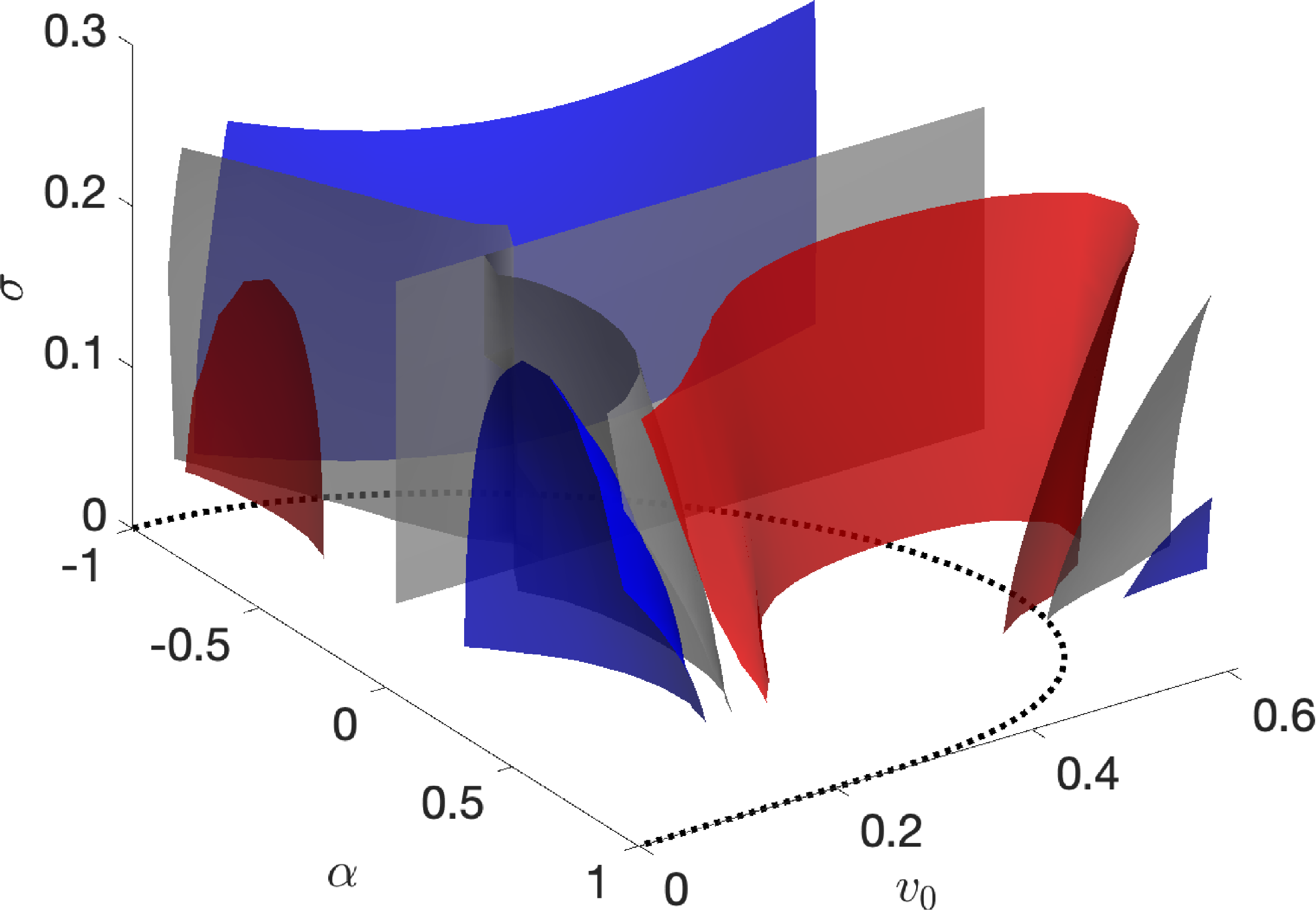}
\caption{Isosurfaces of the depletion index $I_D$ in the $(v_0,\alpha,\sigma)$ parameter space. Blue surfaces: $I_D = -0.15$. Grey surfaces: $I_D = 0$. Red surfaces: $I_D = 0.1$. The black dotted curve in the $(v_0,\alpha)$ plane is the bifurcation curve shown in Fig.~\ref{fig:vBif}.}\label{fig:Id}
\end{figure}
We perform a systematic exploration of the $(v_0,\alpha,\sigma)$ parameter space.
The results are plotted in Fig.~\ref{fig:Id} as isosurfaces of $I_D$.
The transitions from high-shear to low-shear depletion ($I_D <0$ to $I_D > 0$) are indicated by the grey surfaces.
Clearly, these transitions are highly dependent on $v_0$ and $\alpha$, though they only depend weakly on $\sigma$ for the range of $\sigma$ values considered here.
Surprisingly, the transitions appear to be independent of the bifurcation in the swimmer phase space discussed in Sec.~\ref{sec:det}, indicated by the black dotted curve in Fig.~\ref{fig:Id}.

\begin{figure}
\centering
\includegraphics[width=\textwidth]{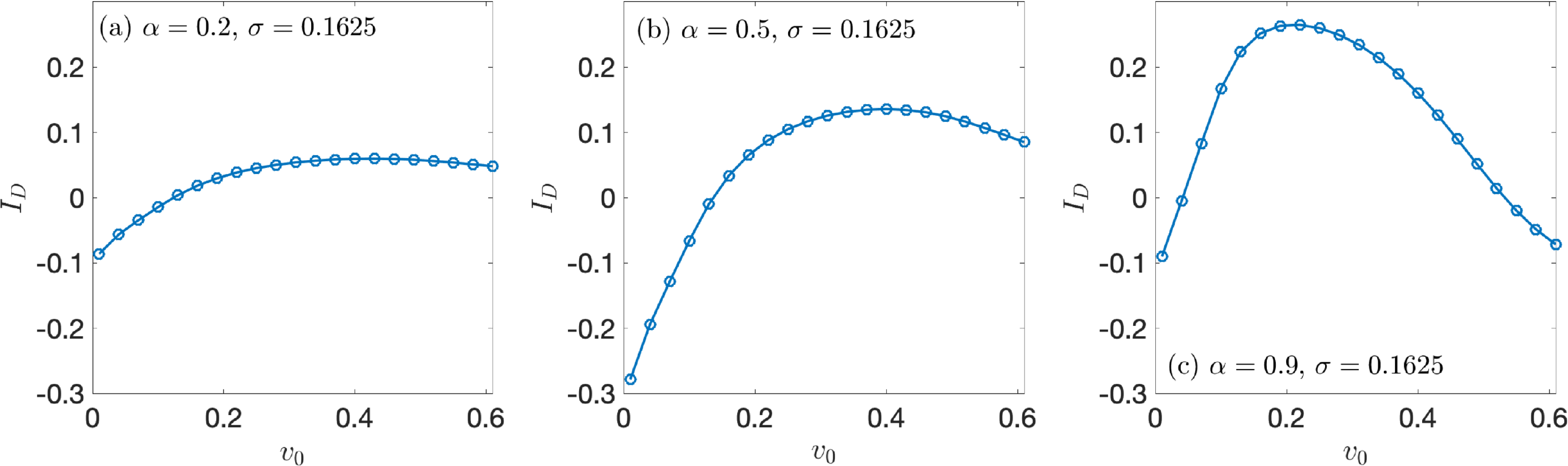}
\caption{Depletion index $I_D$ dependence on $v_0$ for selected values of $\alpha > 0$ and $\sigma = 0.1625$.}\label{fig:Idv0}
\end{figure}
The speed- and shape-dependent depletion behavior of swimmers in the Kolmogorov flow is similar to the experimental observations reported in Refs.~\cite{Rusconi2014,Barry2015} for a channel flow.
These experiments measured $C(y)$ and $I_D$ as a function of maximum flow speed for several swimming microorganisms, with differing shapes and stroke patterns.
We qualitatively mimic this type of experiment by taking cuts of Fig.~\ref{fig:Id} at fixed $\sigma$ and $\alpha$, leading to $I_D(v_0)$ as plotted in Fig.~\ref{fig:Idv0}.
High $v_0$ corresponds to low flow speed, while small $v_0$ corresponds to high flow speed.
Figure \ref{fig:Idv0} shows that in all cases where $\alpha > 0$, the depletion index is maximized at an intermediate $v_0$.
Furthermore, for extremely elongated swimmers such that $\alpha$ is close to $1$, $I_D > 0$ for a wide range of $v_0$ (Fig.~\ref{fig:Idv0}c).
These observations are consistent with experiments on slender swimming bacteria in channel flows, which found that $I_D > 0$ and is maximized for an intermediate flow speed \cite{Rusconi2014}.
On the other hand, for more rounded swimmers for which $\alpha$ is closer to $0$, the depletion index goes from positive to negative as $v_0$ decreases, staying negative over a fairly wide range of $v_0$ (Figs.~\ref{fig:Idv0}a and \ref{fig:Idv0}b).
This is indicative of a transition from low- to high-shear depletion with increasing flow speed, which has been observed in experiments on motile phytoplankton in channel flows \cite{Barry2015}.
The most extreme high-shear depletion with increasing flow speed depends on $\sigma$ and $\alpha$. 
For the value of $\sigma$ used in Fig.~\ref{fig:Idv0}, $I_D(v_0)$ becomes most negative with decreasing $v_0$ for intermediate values of $\alpha$ (i.e.\ intermediate aspect ratios, Fig.~\ref{fig:Idv0}b), while the effect is much more modest when $\alpha$ is close to $0$ (Fig.~\ref{fig:Idv0}a, aspect ratio close to $1$).

\section{Weak-noise behavior via the averaging principle}\label{sec:avg}
In Sec.~\ref{sec:noise}, we showed that rotational diffusion leads to a pronounced, nonuniform probability density of swimmers in phase space.
Nonuniform steady states are typical for dissipative---i.e., non-conservative---dynamical systems perturbed by noise.
In that case, the probability density is peaked around the dissipative system's attractors (e.g.\ stable fixed points or limit cycles) and is shaped by the balance between diffusion and phase space contraction around the attractors.
In our case, Eq.~\eqref{eq:kolmogorov_nondim} is conservative, meaning the phase space does not possess attractors.
Hence, the dynamics of Eq.~\eqref{eq:kolmogorov_nondim} alone cannot explain the regions of phase space where the swimmer density accumulates.
This effect is specifically caused by the interplay between rotational diffusion and conservative dynamics in Eq.~\eqref{eq:kolmogoroVde} \cite{Rusconi2014}.
In this section, we derive a model that captures this interplay by applying an averaging principle for conservative systems that is valid in the weak-noise limit. 
The result is a reduced drift-diffusion model that describes how a swimmer randomly moves across the deterministic orbits of its phase space.
This model allows us to decompose the swimmer's steady-state distribution $P(y,\theta)$ into the product of a probability density on the space of swimmer orbits, parametrized by the deterministic constant of motion $\Psi$, and a kinematic factor accounting for the orientation-dependent angular velocity of elongated particles.

\subsection{Reduced drift-diffusion model}\label{sec:derivation}
To gain insight into the interplay between noise and the conservative dynamics, we calculate the evolution equation for the function $\Psi(y(t),\theta(t))$ under Eq.~\eqref{eq:kolmogoroVde}.
Because $\Psi$ is a function of the stochastic process $(y(t),\theta(t))$, its evolution equation must be derived using It\^o's Lemma \cite{Hassler2016}, which yields the stochastic differential equation
\begin{equation}\label{eq:dPsi}
{\rm d} \Psi = \frac{\sigma^2}{2} \frac{\partial ^2 \Psi}{\partial \theta^2} {\rm d} t + \sigma \frac{\partial \Psi}{\partial \theta} {\rm d} W.
\end{equation}
When $\sigma = 0$, ${\rm d \Psi} = 0$, which implies the conservation of $\Psi$ with no noise, as expected.
Obviously, noise introduces dissipation, in the sense that $\Psi$ is no longer conserved.
When Eq.~\eqref{eq:dPsi} is integrated, we obtain
\begin{equation}\label{eq:Psi_t}
\Psi(t) - \Psi(0) = \frac{\sigma^2}{2} \int_0^t \frac{\partial ^2 \Psi}{\partial \theta^2}(y(t),\theta(t)) {\rm d} t + \sigma \int_0^t \frac{\partial \Psi}{\partial \theta}(y(t),\theta(t)) {\rm d} W.
\end{equation}
Clearly, when $\sigma$ is very small, Eq.~\eqref{eq:Psi_t} implies $\Psi$ changes very slowly.
We accordingly rescale time as $t = \tau/\sigma^2$, under which ${\rm d} W \mapsto \sigma^{-1} {\rm d} W$, so that Eq.~\eqref{eq:Psi_t} becomes
\begin{equation}\label{eq:Psi_tau}
\Psi(\tau) - \Psi(0) = \frac{1}{2} \int_0^\tau \frac{\partial ^2 \Psi}{\partial \theta^2}(y(\tau),\theta(\tau)) {\rm d} \tau +  \int_0^\tau \frac{\partial \Psi}{\partial \theta}(y(\tau),\theta(\tau)) {\rm d} W.
\end{equation}
For sufficiently small $\sigma$, a swimmer will complete many oscillations around a periodic orbit at a fixed $\Psi$ (see Fig.~\ref{fig:det_phase_portrait}) before its $\Psi$ value will have drifted appreciably.
Hence, its motion may be decomposed into fast motion around the deterministic periodic orbits at fixed $\Psi$, and slow motion transverse to the periodic orbits, caused by noise.
In the $\sigma \rightarrow 0$ limit, we can thus approximate the terms on the right-hand side of Eq.~\eqref{eq:Psi_tau} by averaging them over one period of the current orbit at fixed $\Psi$.
This averaging principle is derived rigorously for general two-dimensional conservative dynamical systems perturbed by white noise in Ref.~\cite{Freidlin2012} and references therein.
We describe the technique here and apply it to the swimmer in the Kolmogorov flow.

The averaged equations are as follows.
The assumption that $\Psi$ changes slowly implies that the integrand of the first term of Eq.~\eqref{eq:Psi_tau} can be approximated by its average value over one period of the orbit with fixed $\Psi$, which we denote by $f(\Psi)$.
The function $f$ is given by
\begin{equation}\label{eq:f}
f(\Psi) = \frac{1}{2 T(\Psi)} \oint \frac{\partial^2 \Psi}{\partial \theta^2} {\rm d} \tau,
\end{equation}
where $T(\Psi)$ is the period of this orbit, and
\begin{equation}\label{eq:drift}
\frac{1}{2} \int_0^\tau \frac{\partial ^2 \Psi}{\partial \theta^2}(y(\tau),\theta(\tau)) {\rm d} \tau \approx \int_0^\tau f (\Psi(\tau)) {\rm d} \tau.
\end{equation}
Meanwhile, a standard result from stochastic processes is that the stochastic integral that is the second term of Eq.~\eqref{eq:Psi_tau} is given by 
\begin{equation}\label{eq:stochInt}
\int_0^\tau \frac{\partial \Psi}{\partial \theta} {\rm d} W = W\left( \int_0^\tau \left(\frac{\partial \Psi}{\partial \theta} \right)^2 {\rm d} \tau \right).
\end{equation}
Equation \eqref{eq:stochInt} essentially states that the effect of the prefactor in front of the noise increment ${\rm d} W$ is to rescale the time elapsed along the realization of the Wiener process $W(\tau)$ by the integrated variance of the noise increment, $(\partial \Psi/\partial \theta)^2 {\rm d} \tau$.
The integrand on the right-hand side of Eq.~\eqref{eq:stochInt} can also be approximated using the averaging principle.
We define the averaged diffusivity as
\begin{equation}\label{eq:D}
D(\Psi) = \frac{1}{2 T(\Psi)} \oint \left(\frac{\partial \Psi}{\partial \theta}\right)^2 {\rm d} \tau.
\end{equation}
Thus, the righthand side of  Eq.~\eqref{eq:stochInt} can be approximated by $W(\int_0^\tau 2 D(\Psi(\tau)) {\rm d} \tau)$, and therefore
\begin{equation}\label{eq:diffusion}
\int_0^\tau \frac{\partial \Psi}{\partial \theta} {\rm d} W \approx \int_0^\tau \sqrt{2 D(\Psi(\tau))} {\rm d} W.
\end{equation}
Substituting Eqs.~\eqref{eq:drift} and \eqref{eq:diffusion} into Eq.~\eqref{eq:Psi_tau}, we obtain the approximate drift-diffusion process after averaging,
\begin{equation}\label{eq:PsiDrift}
\Psi(\tau) - \Psi(0) \approx \int_0^\tau f(\Psi(\tau)) {\rm d} \tau + \int_0^{\tau} \sqrt{2 D(\Psi(\tau))} {\rm d}W.
\end{equation}
By time-averaging, we have reduced the two-dimensional drift-diffusion process \eqref{eq:kolmogoroVde} to the one-dimensional process \eqref{eq:PsiDrift}, which describes how a swimmer diffuses across the deterministic orbits. 

\begin{figure}
\centering
\includegraphics[width=0.9\textwidth]{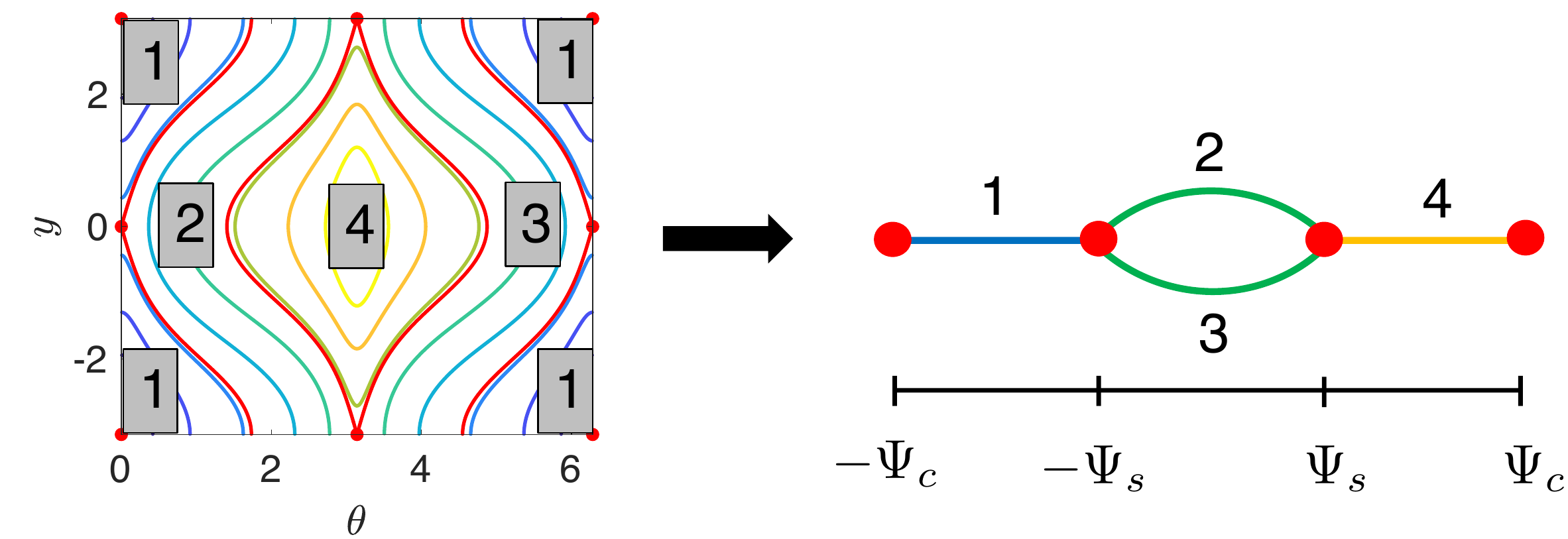}
\caption{Reduction of the swimmer phase space to a graph. Each edge of the graph corresponds to the labeled region of phase space, and the nodes correspond to fixed points and separatrices at the boundaries of each region. $\Psi$ is a coordinate along the edges of the graph.}\label{fig:graph}
\end{figure}
Equation \eqref{eq:PsiDrift} is valid in each of the topologically distinct regions of phase space, i.e.\ the islands and the jets.
However, the drift-diffusion processes in each of these regions must be stitched together with proper boundary conditions in order to describe the averaged dynamics on the full phase space \cite{Freidlin2012}.
After averaging, the $(y,\theta)$ phase space can be reduced to a graph, depicted in Fig.~\ref{fig:graph}.
Each edge represents one of the topologically distinct regions of phase space, i.e. an island or a jet.
Each point on the interior of an edge represents a distinct periodic orbit with a particular value of  $\Psi$.
Thus, $\Psi$ is a coordinate along each of the edges.
We denote the absolute value of $\Psi$ on the separatrix as $\Psi_s = |\Psi(0,0)|$ and the maximum value of $\Psi$, occurring at the center fixed point $(y,\theta) = (0,\pi)$, as $\Psi_c = \Psi(0,\pi)$.
By symmetry, the separatrices occur at $\Psi = \pm \Psi_s$ and the centers occur at $\Psi = \pm \Psi_c$.
Thus, the nodes of the graph in Fig.~\ref{fig:graph} at $\Psi = \pm \Psi_c$ represent the centers, and the nodes at $\Psi = \pm \Psi_s$ represent the saddles and separatrices that are on the island-jet boundaries.
The graph in Fig.~\ref{fig:graph} has the same structure regardless of whether the jets exhibit cross-stream drift or rotational drift.

To each edge, we associate a time-dependent probability density $p_i(\Psi,\tau)$, with $i \in \{1,2,3,4 \}$.
The densities $p_i$ evolve according to the Fokker-Planck equations associated with Eq.~\eqref{eq:PsiDrift},
\begin{align} \label{eq:PsiFPs}
\frac{\partial p_i}{\partial \tau} = -\frac{\partial }{\partial \Psi} \left( f_i p_i \right) + \frac{\partial^2}{\partial\Psi^2} \left(D_i p_i \right) = -\frac{\partial J_i}{\partial\Psi},
\end{align}
where $J_i$ is the probability current density
\begin{equation}\label{eq:j}
J_i = f_i p_i - \frac{\partial}{\partial \Psi} \left(D_i p_i \right).
\end{equation}
In Eqs.~\eqref{eq:PsiFPs} and \eqref{eq:j}, $f_i$ and $D_i$ are the averaged drifts and diffusions [Eqs.~\eqref{eq:f} and \eqref{eq:D} respectively] evaluated in the regions of phase space corresponding to the edges of the graph in Fig.~\ref{fig:graph}.
At the nodes of the graph where the islands and jets meet, the local conservation of probability implies that the total probability current density entering a node must equal the total probability current density leaving the node, similar to Kirchoff's first law for circuits.
To obtain the total probability current density entering (leaving) a node, one sums over the $J_i$ for edges $i$ connected to that node such that the node is approached in the direction of increasing (decreasing) $\Psi$.
The boundary conditions are thus
\begin{subequations}\label{eq:BCs}
\begin{align}
J_1(-\Psi_s) & = J_2(-\Psi_s) + J_3(-\Psi_s), \\
J_4(\Psi_s) & = J_2(\Psi_s) + J_3(\Psi_s).
\end{align}
\end{subequations}
Equations \eqref{eq:PsiFPs} and \eqref{eq:BCs} thus constitute a drift-diffusion process on the graph illustrated in Fig.~\ref{fig:graph}, which captures the swimmer dynamics in the $\sigma \rightarrow 0$ limit.

In order to investigate the steady-state behavior of the swimmer in the weak-noise limit, we seek the steady-state solution of Eqs.~\eqref{eq:PsiFPs} and \eqref{eq:BCs}.
Because the jets are identical to each other by symmetry, $f_2=f_3$ and $D_2=D_3$ in Eq.~\eqref{eq:PsiFPs}, and therefore $p_2 = p_3$ must be satisfied in the steady-state.
This allows us to merge the distinct $p_i$ into a single steady-state density, $p_0(\Psi)$, defined over the entire range $-\Psi_c \leq \Psi \leq\Psi_c$, which satisfies
\begin{equation}\label{eq:p0graph}
p_0(\Psi) = \begin{cases}
p_1 & {\rm for}\,\, -\Psi_c \leq \Psi < -\Psi_s, \\
2 p_2 & {\rm for}\,\, -\Psi_s \leq \Psi <\Psi_s, \\
p_4 & {\rm for} \,\, \Psi_s \leq \Psi \leq \Psi_c.
\end{cases}
\end{equation}
The density $p_0$ satisfies the steady-state Fokker-Planck equation
\begin{equation}\label{eq:PsiFP}
-\frac{\rm d}{{\rm d}\Psi} \left( f p_0 \right) + \frac{{\rm d}^2}{{\rm d} \Psi^2} \left(D p_0 \right) = 0.
\end{equation}
Equation \eqref{eq:PsiFP} can be integrated once, which after rearrangement gives
\begin{equation}\label{eq:PsiFP1}
p_0' = \frac{1}{D}\left[ \left(f - D' \right) p_0 + C_1 \right],
\end{equation}
where $C_1$ is an integration constant and $(\cdot)' \equiv {\rm d} (\cdot)/{\rm d} \Psi$.
Equation \eqref{eq:PsiFP1} is linear, and its solution is
\begin{equation}\label{eq:PsiFPSol1}
p_0(\Psi) =\exp \left[ \int_{-\Psi_c}^\Psi \frac{f(\Psi_1) - D'(\Psi_1)}{D(\Psi_1)} {\rm d} \Psi_1 \right] \left\{ C_0  + C_1 \int_{-\Psi_c}^\Psi \frac{\exp \left[ - \int_{-\Psi_c}^{\Psi_1}  \frac{f(\Psi_2) - D'(\Psi_2)}{D(\Psi_2)} {\rm d} \Psi_2 \right]}{D(\Psi_1)} {\rm d} \Psi_1 \right\},
\end{equation}
where $C_0$ is another integration constant.
Due to the symmetry of the problem, the conditions
\begin{subequations}\label{eq:conditions}
\begin{align}\label{eq:cond1}
 & p_0(-\Psi_c) = p_0(\Psi_c), \\ \label{eq:cond2}
 & f(\Psi) = -f(-\Psi), \\ \label{eq:cond3}
 & D(\Psi) = D(-\Psi) 
\end{align}
\end{subequations}
must be satisfied.
We substitute Eq.~\eqref{eq:PsiFPSol1} into Eq.~\eqref{eq:cond1}, which gives
\begin{equation}\label{eq:symmetrycond}
C_0 = C_0 + C_1 \int_{-\Psi_c}^{\Psi_c} \frac{\exp \left[ - \int_{-\Psi_c}^{\Psi_1}  \frac{f(\Psi_2) - D'(\Psi_2)}{D(\Psi_2)} {\rm d} \Psi_2 \right]}{D(\Psi_1)} {\rm d} \Psi_1,
\end{equation}
where we have used Eqs.~\eqref{eq:cond2} and \eqref{eq:cond3} in evaluating the first term of Eq.~\eqref{eq:PsiFPSol1} on the right-hand side of Eq.~\eqref{eq:cond1}.
Equation \eqref{eq:symmetrycond} implies $C_1 = 0$.
Hence, the steady-state $\Psi$ distribution has the form
\begin{equation}\label{eq:PsiDist}
p_0(\Psi) = \frac{C_0}{D(\Psi)} \exp \left[ \int_{-\Psi_c}^\Psi \frac{f(\Psi_1)}{D(\Psi_1)} {\rm d} \Psi_1 \right],
\end{equation}
where we have absorbed another constant into $C_0$, which now plays the role of normalization constant.
Equation \eqref{eq:PsiDist} is the invariant density of the reduced drift-diffusion model \eqref{eq:PsiDrift}.
Physically, it is the steady-state probability density of finding a swimmer on a deterministic trajectory with a particular constant of motion $\Psi$ in the $\sigma \rightarrow 0$ limit.

We now use $p_0$ to reconstruct the phase-space distribution $P_0(y,\theta)$ that is obtained in the $\sigma \rightarrow 0$ limit.
The idea is to make a change of coordinates $(y,\theta) \mapsto (\Psi,s)$, where $\Psi = \Psi(y,\theta)$ is defined by Eq.~\eqref{eq:Psi} and $s = s(y,\theta)$ is the elapsed Euclidean arclength along a trajectory at fixed $\Psi$, defined by
\begin{equation}\label{eq:ds}
{\rm d} s = \sqrt{{\rm d} y^2 + {\rm d}\theta^2} = |\dot{\bq}| {\rm d} t,
\end{equation}
where $\bq = (y,\theta)$ and $|\dot{\bq}| \equiv (\dot{y}^2 + \dot{\theta}^2)^{1/2}$.
The coordinates $(\Psi,s)$ are akin to action-angle variables in Hamiltonian mechanics.
Note that this is a local rather than global change of coordinates, because for each value of $\Psi$ in the jet region, there are two distinct orbits, one above and one below the separatrix (see Fig.~\ref{fig:graph}).
Under this change of coordinates, the phase-space probability distribution must transform as
\begin{align}
P_0(y,\theta) {\rm d} y {\rm d} \theta & = P^*(\Psi,s) {\rm d} \Psi {\rm d} s \nonumber \\ \label{eq:changeofco}
& =  P^*(\Psi,s) \left| \det \left[ \frac{\partial (\Psi, s) }{\partial (y,\theta) } \right] \right| {\rm d} y {\rm d} \theta.
\end{align}
Here, $P^*(\Psi,s)$ is the steady-state distribution in $(\Psi,s)$ coordinates.
Under the assumptions of the averaging principle, the fast motion along the $s$ coordinate is decoupled from the slow motion along the $\Psi$ coordinate, so the joint density $P^*(\Psi,s)$ must be the product of the density of the $\Psi$ coordinate, $p_0(\Psi)$, and the invariant density along an orbit of fixed $\Psi$.
The latter is inversely proportional to the phase-space speed $|\dot{\bq}|$ at each point.
In addition, we must account for the fact that for values of $\Psi$ in the jet region, the two jets share probability equally [see Eq.~\eqref{eq:p0graph}].
Therefore, we obtain
\begin{equation}\label{eq:pstar}
P^*(\Psi,s) = \begin{cases}
\frac{p_0(\Psi)}{T(\Psi) |\dot{\bq}|} & {\rm for}\,\,\, |\Psi| > \Psi_s, \\
\frac{1}{2} \frac{p_0(\Psi)}{T(\Psi) |\dot{\bq}|} & {\rm otherwise}.
\end{cases}
\end{equation}
The orbit period $T$ is included in Eq.~\eqref{eq:pstar} to normalize the invariant density of the $s$ coordinate.

The calculation of the determinant in Eq.~\eqref{eq:changeofco} requires the partial derivatives of $\Psi$ and $s$ with respect to $(y,\theta)$.
The partial derivatives of $\Psi$ are straightforward to obtain from Eq.~\eqref{eq:Psi}.
The partial derivatives of $s$ follow from Eq.~\eqref{eq:ds},
\begin{equation}\label{eq:sderiv}
\frac{{\rm d} s}{{\rm d} t	} = |\dot{\bq}| = \nabla s \cdot \dot{\bq},
\end{equation}
where $\nabla \equiv (\partial/\partial y, \partial/\partial \theta)$.
From Eq.~\eqref{eq:sderiv}, it follows that
\begin{equation}
\nabla s = \frac{\dot{\bq}}{|\dot{\bq}|}.
\end{equation}
Thus, a straightforward calculation leads to
\begin{equation}\label{eq:det}
\det \left[\frac{\partial (\Psi, s) }{\partial (y,\theta) } \right] = -\frac{2 |\dot{\bq}|}{1 - \alpha \cos 2\theta}.
\end{equation}
Combining Eqs.~\eqref{eq:changeofco}, \eqref{eq:pstar}, and \eqref{eq:det}, we finally obtain
\begin{equation}\label{eq:p0yth}
P_0(y,\theta) = \frac{g(\Psi(y,\theta))}{1-\alpha \cos 2\theta},
\end{equation}
where
\begin{equation}\label{eq:g}
g(\Psi) = \begin{cases}
\frac{2 p_0(\Psi)}{T(\Psi)} & {\rm for}\,\,\, |\Psi| > \Psi_s, \\
\frac{p_0(\Psi)}{T(\Psi)} & {\rm otherwise}.
\end{cases}
\end{equation}

Equation \eqref{eq:p0yth}  shows that the nonuniformity of the steady-state probability density $P(y,\theta)$ in the limit of small noise results from two effects: rotational slowdown and noise-induced drift.
Regardless of where one is in phase space, the density is modulated by the factor of $(1 - \alpha \cos 2\theta)^{-1}$.
This factor accounts for the slowdown of rotating elongated particles, caused by the orientation-dependent angular velocity in  Eq.~\eqref{eq:thdot}.
In addition to this, the density is modulated by the function $g(\Psi)$, which weights each point $(y,\theta)$ according to the deterministic orbit it belongs to, indexed by $\Psi(y,\theta)$.
The modulation by $g$ accounts for the slow dynamics across the deterministic periodic orbits, encapsulated by the averaged drift-diffusion model \eqref{eq:PsiDrift}.

\subsection{Unraveling density variations and depletion}
\begin{figure}
\centering
\includegraphics[width=\textwidth]{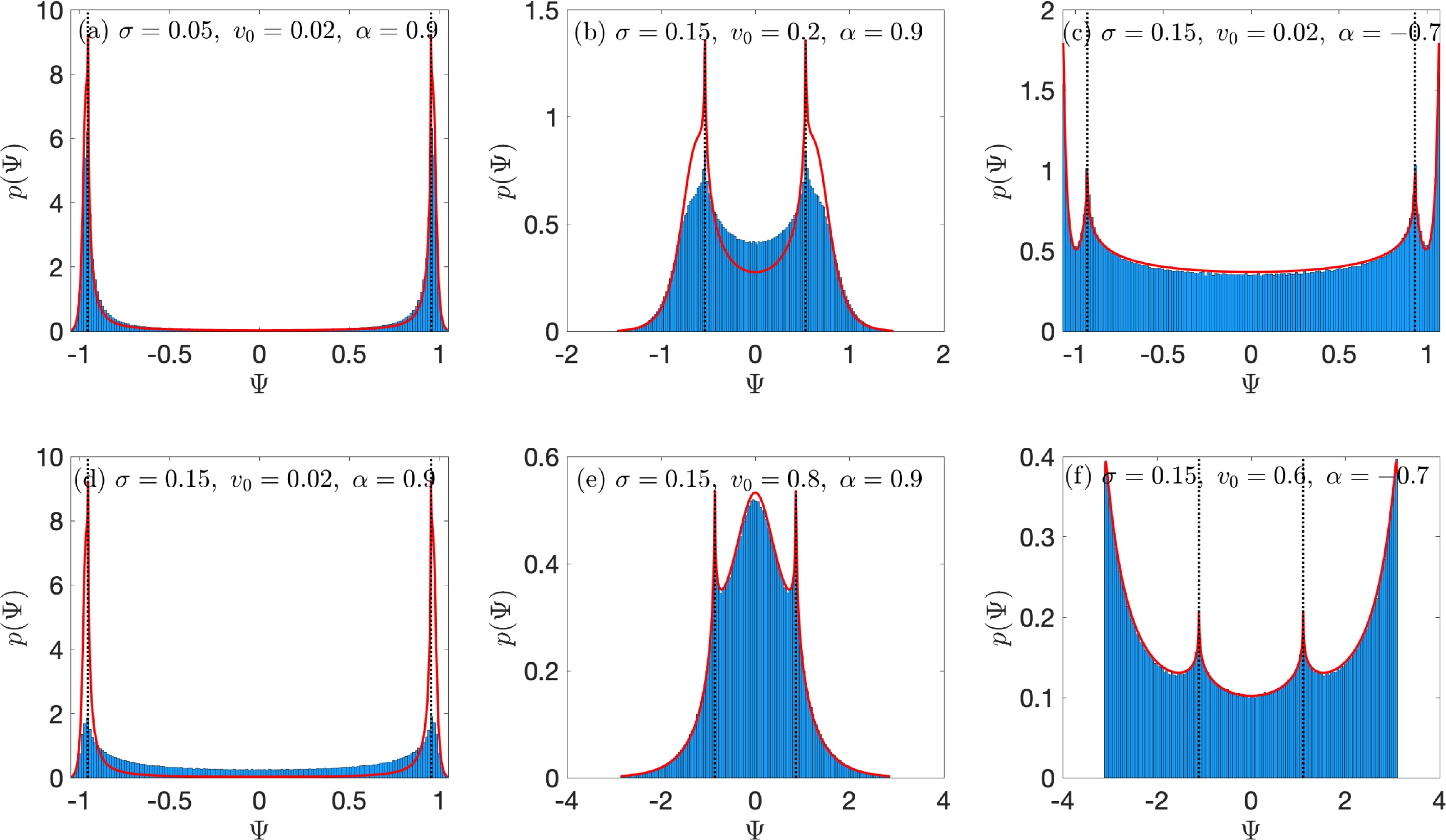}
\caption{Comparison of the steady-state $\Psi$ distributions from the averaging principle [$p_0(\Psi)$, red curves] and those obtained from solving the Fokker-Planck equation \eqref{eq:fp} and applying Eq.~\eqref{eq:pPsi}. The parameters are the same as in Fig.~\ref{fig:Pyth}. Black dotted lines indicate the $\Psi$ values of the separatrices $\pm \Psi_s$. Note that $p_0(\Psi)$ does not depend on $\sigma$, so the curves plotted in panels (a) and (d) are the same.}\label{fig:psiDistAlpha1}
\end{figure}
We proceed by evaluating $p_0(\Psi)$ through the numerical evaluation of Eqs.~\eqref{eq:f}, \eqref{eq:D}, and \eqref{eq:PsiDist} for selected parameters $(v_0,\alpha)$ \cite{code}.
The resulting distributions are compared against the distributions $p(\Psi)$ obtained directly from the numerical solutions $P(y,\theta)$ to the Fokker-Planck equation \eqref{eq:fp}.
We compute $p(\Psi)$ using
\begin{equation}\label{eq:pPsi}
p(\Psi) \approx \frac{1}{\Delta \Psi} \int_{\Psi}^{\Psi+\Delta \Psi} P(y,\theta) {\rm d} y {\rm d} \theta
\end{equation}
for small $\Delta \Psi$.
Example results are presented in Fig.~\ref{fig:psiDistAlpha1}.
Note that in each panel of Fig.~\ref{fig:psiDistAlpha1}, $p$ is plotted for the range $-\Psi_c \leq \Psi \leq \Psi_c$.
Because $\Psi_c$ depends on $v_0$ and $\alpha$, the numerical ranges of $p$ are different in each panel.
For sufficiently small $\sigma$, we see excellent agreement between $p_0$ predicted by the averaging principle and $p$ obtained from the exact solutions of the Fokker-Planck equation (Figs.~\ref{fig:psiDistAlpha1}a, \ref{fig:psiDistAlpha1}c, \ref{fig:psiDistAlpha1}e, and \ref{fig:psiDistAlpha1}f).
As the intensity of noise increases, the two $\Psi$ distributions begin to deviate from each other, as seen when comparing Figs.~\ref{fig:psiDistAlpha1}a and \ref{fig:psiDistAlpha1}d.
The assumption underlying the averaging principle is that the noise-induced drift across orbits is slow compared to the fast motion around the deterministic periodic orbits.
We estimate the time scale of the fast motion using the period $T(\Psi_c)$ of small oscillations around the centers $(y,\theta) = (0,\pi)$ and $(\pi,0)$, which is given by $T(\Psi_c) = 2^{3/2}\pi/\sqrt{v_0(1-\alpha)}$.
Meanwhile, the time scale of the noise-induced drift can be estimated as $\tau_{\rm noise} = 1/\sigma^2$ [see Eq.~\eqref{eq:Psi_t}].
Hence, we expect good agreement between $p_0$ and $p$ when $\tau_{\rm noise} \gg T(\Psi_c)$, which can be rearranged to give
\begin{equation}\label{eq:valid1}
\sigma^2 \ll \frac{\sqrt{v_0(1-\alpha)}}{2^{3/2}\pi}.
\end{equation}
Equation \eqref{eq:valid1} can also be expressed in terms of the dimensional quantities as
\begin{equation}
\frac{D_R w}{\sqrt{VU(1-\alpha)}} \ll 1,
\end{equation}
where we have dropped the numerical constants.
The scaling behavior in Eq.~\eqref{eq:valid1} explains why for the same value of $\sigma = 0.15$, the agreement between $p_0$ and $p$ is better for large $v_0$ (Fig.~\ref{fig:psiDistAlpha1}e compared to Figs.~\ref{fig:psiDistAlpha1}b and \ref{fig:psiDistAlpha1}d) and $\alpha < 0$ (Fig.~\ref{fig:psiDistAlpha1}c compared to Fig.~\ref{fig:psiDistAlpha1}d).

\begin{figure}
\centering
\includegraphics[width=\textwidth]{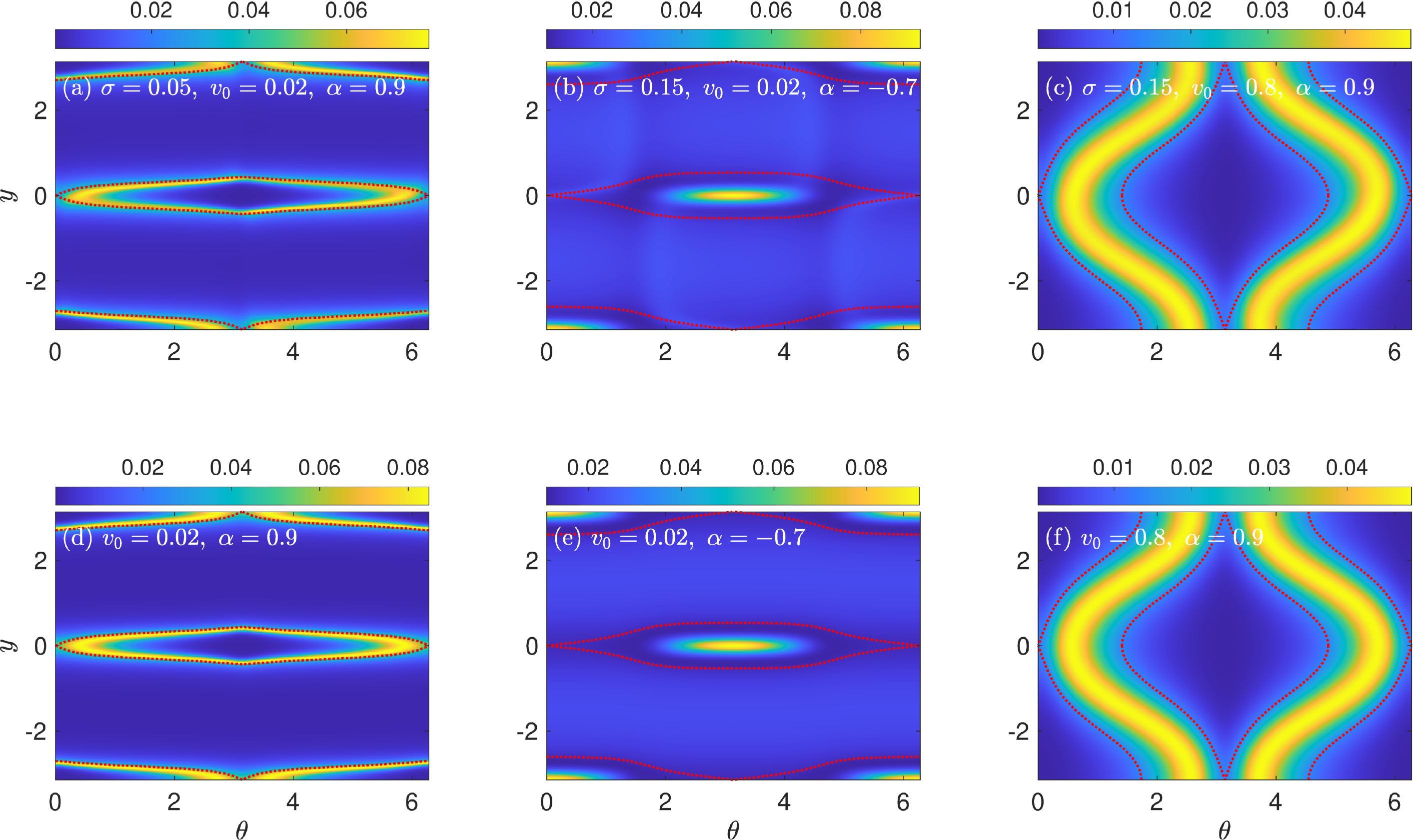}
\caption{Comparison of (a--c) the phase-space distributions with the rotational slowdown effect filtered out, $P(y,\theta)(1-\alpha \cos 2\theta)$,  with (d--f) the distributions $P_0(y,\theta)(1-\alpha \cos 2\theta)$ predicted by the averaging model given in Eq.~\eqref{eq:p0yth}.}\label{fig:averagingComp}
\end{figure}
Next, we compare the probability densities $P_0$ predicted by the reduced model and given by Eq.~\eqref{eq:p0yth} to the densities $P$ obtained by solving the Fokker-Planck equation \eqref{eq:fp}.
In particular, we focus on assessing the extent to which the non-uniformity of $P$ is caused by noise-induced drift, as opposed to rotational slowdown.
Owing to the form of $P$ in the $\sigma \rightarrow 0$ limit given in Eq.~\eqref{eq:p0yth}, we filter out the effect of rotational slowdown by multiplying the densities by $1-\alpha \cos(2\theta)$.
The results are shown in Fig.~\ref{fig:averagingComp} for selected parameters.
Figures \ref{fig:averagingComp}a, \ref{fig:averagingComp}b, and \ref{fig:averagingComp}c show Fokker-Planck densities $P$ with the rotational slowdown modulation removed, and they correspond to Figs.~\ref{fig:Pyth}a, \ref{fig:Pyth}c, and \ref{fig:Pyth}e, respectively.
For $\alpha > 0$ (Figs.~\ref{fig:averagingComp}a and \ref{fig:averagingComp}c), the peaks at $\theta = 0$ and $\pi$ are gone, confirming that these peaks are caused by rotational slowdown.
Similarly, for $\alpha < 0$ (Fig.~\ref{fig:averagingComp}b), the peaks at $\theta = \pi/2$ and $3\pi/2$ are gone.
Furthermore, the filtered densities $P(1-\alpha \cos 2\theta)$ from the Fokker-Planck simulations (Figs.~\ref{fig:averagingComp}a--c) agree well with the corresponding filtered densities $P_0(1-\alpha \cos 2\theta)$ obtained using the averaging technique in the weak-noise limit (Figs.~\ref{fig:averagingComp}d--f).
After filtering out the rotational slowdown, all that remains in $P_0$ is $g(\Psi(y,\theta))$.
Thus,  Figs.~\ref{fig:averagingComp}d--f display the orbit weight-factor $g$ at each point of phase space.

\begin{figure}
\centering
\includegraphics[width=0.9\textwidth]{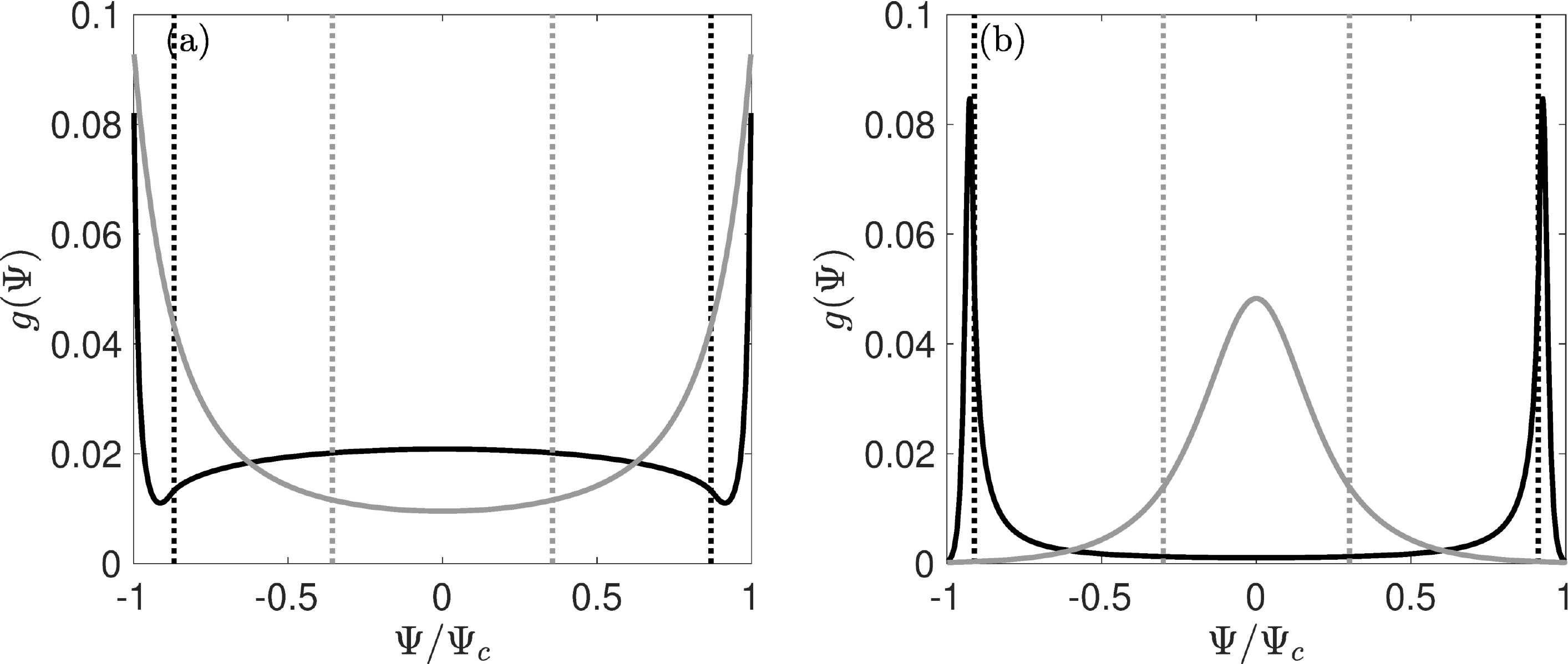}
\caption{Phase space weight factors $g$ for (a) $\alpha = -0.7$ and (b) $\alpha = 0.9$. The solid curves are $g(\Psi)$, and the dotted curves represent the separatrices at $\pm \Psi_s$. Different shades represent different swimming speeds $v_0$. (a) Black is $v_0 = 0.02$, and grey is $v_0 = 0.6$. (b) Black is $v_0 = 0.02$, and grey is $v_0 = 0.8$.}\label{fig:gFunc}
\end{figure}
Figure \ref{fig:gFunc} shows $g(\Psi)$ for several representative combinations of $v_0$ and $\alpha$.
Note that $g(\Psi)$ appears to be a smooth function, while $p_0(\Psi)$ possesses singularities at the separatrices $\Psi = \pm \Psi_s$, as seen in Fig.~\ref{fig:psiDistAlpha1}.
It can be shown that these singularities occur because in Eq.~\eqref{eq:PsiDist}, $D(\pm \Psi_s) = 0$.
However, from Eq.~\eqref{eq:g} we have $g \propto p_0/T$, and $T$ also diverges at the separatrices.
Evidently, these singularities exactly cancel each other out, making $g$ a smooth function.

The good agreement between Figs.~\ref{fig:averagingComp}a--c and \ref{fig:averagingComp}d--f both validates the reduced model derived in Sec.~\ref{sec:derivation} and shows the importance of noise-induced drift in shaping the overall phase-space densities.
In particular, the shape of $g(\Psi)$ for different values of swimmer speed $v_0$ and shape $\alpha$ explains many of our observations concerning the steady-state distributions $P(y,\theta)$ in Sec.~\ref{sec:noise}.
For example, for $\alpha < 0$, $g$ is always peaked at the boundaries $\pm \Psi_c$ (Fig.~\ref{fig:gFunc}a).
In phase space, these features manifest as peaks on the centers, as seen in Figs.~\ref{fig:averagingComp}b and \ref{fig:averagingComp}e, as well as Figs.~\ref{fig:Pyth}c and \ref{fig:Pyth}f.
When $v_0$ is sufficiently small, $g$ also has a secondary, broad peak centered on the jet region ($\Psi = 0$), as evidenced by the black curve ($v_0 = 0.02$) in Fig.~\ref{fig:gFunc}a.
This leads to a significant density in the jet regions, as seen in Fig.~\ref{fig:Pyth}c.
As $v_0$ increases, the height of this secondary peak relative to the peaks at $\pm \Psi_c$ decreases, while the local minima on either side of the central peak creep inward.
For sufficiently large $v_0$, these local minima merge and the secondary peak disappears entirely, as evidenced by the grey curve ($v_0 = 0.6$) in Fig.~\ref{fig:gFunc}a.
Thus, as $v_0$ grows, the orbits in the jet region become increasingly suppressed for $\alpha < 0$ swimmers (Fig.~\ref{fig:Pyth}f).

We observe qualitatively different behavior for $g$ when $\alpha > 0$ (Fig.~\ref{fig:gFunc}b).
When $v_0$ is small, $g$ is always peaked near the separatrices, as illustrated by the black curve ($v_0 = 0.02$) in Fig.~\ref{fig:gFunc}b.
Note that the peaks are slightly offset from the separatrices, shifted towards the islands.
This feature explains the bright bands just inside the separatrices seen in the filtered $P(y,\theta)$ distributions in Figs.~\ref{fig:averagingComp}a and \ref{fig:averagingComp}d and the raw distributions in Figs.~\ref{fig:Pyth}a, \ref{fig:Pyth}b, and \ref{fig:Pyth}d.
As $v_0$ increases, both the separatrices and the peaks of $g$ move inward towards $\Psi = 0$.
For sufficiently large $v_0$, these peaks merge and $g$ exhibits a single peak at $\Psi = 0$, as illustrated by the grey curve ($v_0 = 0.8$) in Fig.~\ref{fig:gFunc}b.
Hence, at sufficiently high $v_0$, the orbits in the jet region are weighted the highest, as seen in Figs.~\ref{fig:averagingComp}c and \ref{fig:averagingComp}f.

\begin{figure}
\includegraphics[width=\textwidth]{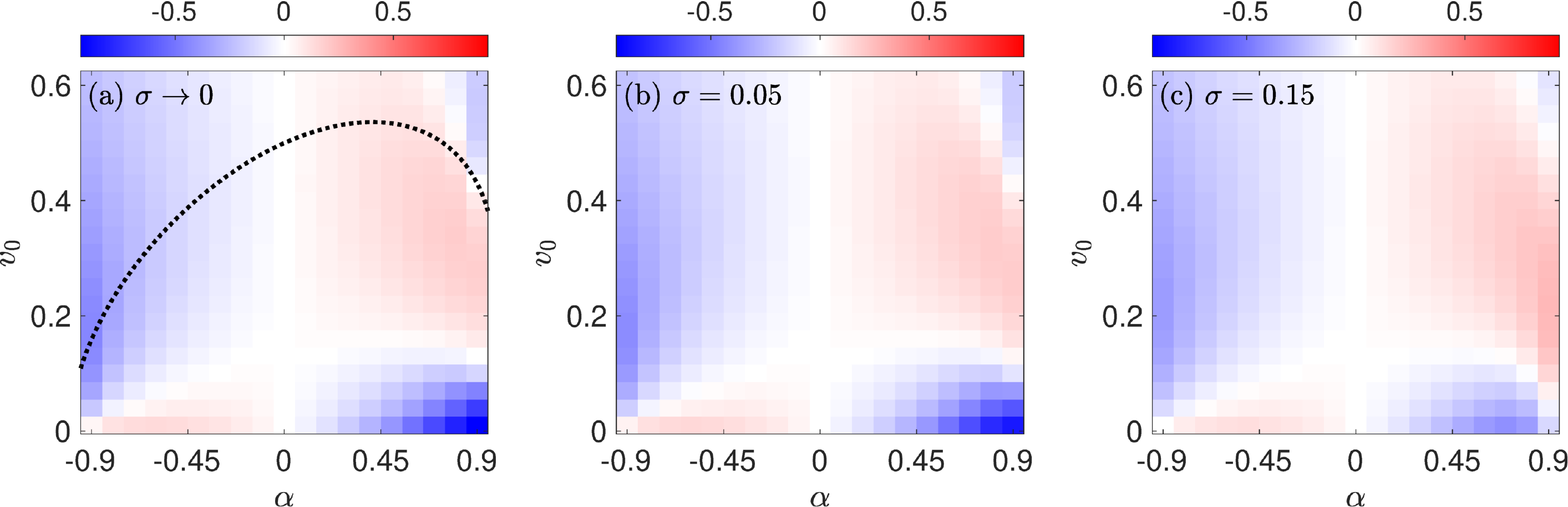}
\caption{Depletion index $I_D$ as a function of $(v_0,\alpha)$ for fixed values of $\sigma$. (a) $\sigma \rightarrow 0$, using the averaged drift-diffusion model and Eq.~\eqref{eq:p0yth}. The black dotted curve is the bifurcation curve from Fig.~\ref{fig:vBif}. (b) $\sigma = 0.05$, from the Fokker-Planck model. (c) $\sigma = 0.15$, from the Fokker-Planck model.}\label{fig:IdComp}
\end{figure}
The variation of $g$ with the swimmer parameters also clarifies the variation of the depletion index $I_D$ as a function of $v_0$ and $\alpha$.
Figure \ref{fig:IdComp}a shows the variation of $I_D$ predicted using $P_0$ given by Eq.~\eqref{eq:p0yth}.
The quantitative agreement with the depletion index from the Fokker-Planck model is excellent for $\sigma = 0.05$ (Fig.~\ref{fig:IdComp}b) and reasonable for $\sigma = 0.15$ (Fig.~\ref{fig:IdComp}c).
For $\alpha < 0$, we can now ascribe the transition from low- to high-shear depletion (positive to negative $I_D$) to the increasing dominance of the peaks of $g$ at $\Psi = \pm \Psi_c$ as $v_0$ increases.
Meanwhile, for $\alpha > 0$, we ascribe the transition from high- to low-shear depletion to the peaks of $g$ tracking the separatrices, which narrowly hug the centerlines for small $v_0$ and gradually widen as $v_0$ increases.

\begin{figure}
\centering
\includegraphics[width=0.8\textwidth]{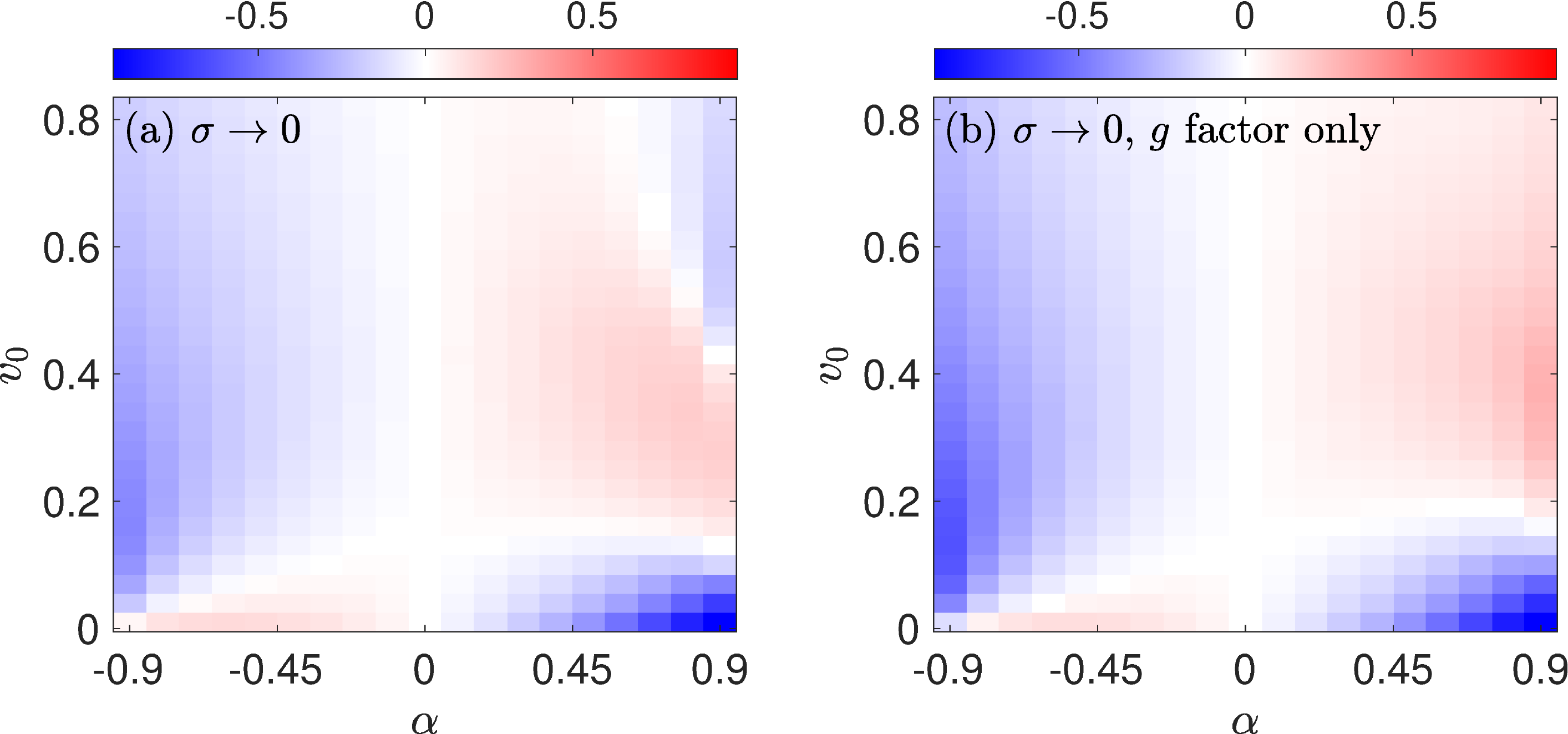}
\caption{Comparison of calculations of the depletion index $I_D$ (a) using $P_0(y,\theta)$ vs. (b) using the rotational slowdown filtered $P_0(1-\alpha \cos 2\theta) = g$.}\label{fig:Idg}
\end{figure}
As for the second transition to high-shear depletion that occurs for $\alpha > 0$ as $v_0$ increases further, i.e.\ the blue region in the upper-right corner of Fig.~\ref{fig:IdComp}a, this is actually due to the interplay between noise-induced drift and rotational slowdown.
To demonstrate this, in Fig.~\ref{fig:Idg} we compare $I_D$ calculated using the full $P_0$ given by Eq.~\eqref{eq:p0yth} (Fig.~\ref{fig:Idg}a) to a calculation of $I_D$ where we use instead the filtered density $P_0(y,\theta)(1-\alpha\cos 2\theta) = g(\Psi(y,\theta))$, which must be normalized (Fig.~\ref{fig:Idg}b).
The two calculations agree well everywhere except for $\alpha$ near $1$ for high $v_0$.
In that region, $I_D$ decreases as $v_0$ increases in both cases, but with rotational slowdown included (Fig.~\ref{fig:Idg}a), the decline of $I_D$ is enhanced to the point that it becomes negative.
For $\alpha$ near $1$ and high $v_0$, $g$ is peaked at $\Psi = 0$ (Fig.~\ref{fig:gFunc}b), so that orbits in the jet region are highly weighted (Figs.~\ref{fig:averagingComp}c and \ref{fig:averagingComp}f).
At the same time, the rotational slowdown factor $1-\alpha \cos 2\theta$ is accentuated, due to a larger $|\alpha|$.
The combination of these two effects leads to a very high concentration of $P(y,\theta)$ near the saddles, modest ridges in the jet regions, and a steep drop-off everywhere else, as seen in Fig.~\ref{fig:Pyth}e.
These elements combined then lead to $I_D < 0$.
Overall, we conclude that noise-induced drift is the dominant effect underlying the $(v_0,\alpha)$ parameter dependence of $I_D$, while rotational slowdown is a secondary effect that becomes most important for fast and highly elongated $\alpha > 0$ swimmers. 

\section{Conclusion}\label{sec:concl}
To conclude, we have analyzed the distribution of microswimmers in the planar, laminar Kolmogorov flow.
These distributions are highly nonuniform for elongated swimmers, and the swimmer concentration may be depleted from either low- or high-shear regions of the flow \cite{Rusconi2014,Barry2015}.
To explain this effect, we derived a reduced model using an averaging technique \cite{Freidlin2012}, which captures the slow motion of swimmers transverse to the deterministic orbits that is induced by weak diffusion.
The model leads to an invariant density on the space of deterministic swimmer orbits, parametrized by the constant of motion $\Psi$ of the deterministic dynamics.
The steady-state phase-space densities predicted by the reduced model are in good agreement with those obtained from the original model, showing that the main cause of depletion is the noise-induced drift of swimmers in phase space.

The averaging technique we employed here can be applied to other problems involving noisy self-propelled particles in fluid flows.
For systems with an effectively two-dimensional phase space, the technique is applicable so long as the system possesses a conserved quantity and the noise is diffusive.
For example, it may also be applied to gyrotactic swimmers in a planar laminar Kolmogorov flow \cite{Santamaria2014} and swimmers with both translational and rotational diffusion \cite{Berman2021b,Thiffeault2021}.
For higher dimensional systems possessing one or multiple conserved quantities, such as swimmers in three-dimensional channel flows \cite{Zottl2013}, a generalization of the two-dimensional theory exists for extracting the slow dynamics induced by weak noise \cite{Freidlin2012}.

Another potentially interesting application would be to understanding the effects of chemotaxis on density variations of swimming microorganisms in fluid flows \cite{Rusconi2014,Locsei2009,Bearon2015}.
Our model can be modified by making the noise intensity $\sigma$ vary with $y$ and $\theta$, in order to mimic the position- and orientation-dependent tumbling rate of chemotactic swimming bacteria in response to a chemical gradient in the flow.
This would modify the averaged drift and diffusion functions $f$ and $D$, defined in Eqs.~\eqref{eq:f} and \eqref{eq:D}, and it would potentially break the symmetry between the two distinct jet regions.
However, the rest of the procedure carried out in Sec.~\ref{sec:avg} would still be applicable.
It is an open question whether an averaging approach would be applicable to models where the rotational noise of the swimmer contains a tumbling component in addition to (or instead of) rotational diffusion \cite{Berman2021b,Locsei2009,Bearon2015}.

\appendix
\section{Numerical method for solution of the Fokker Planck equation}\label{sec:app}
We solve Eq.~\eqref{eq:fp} using Fourier transforms.
We approximate $P(y,\theta)$ as the truncated Fourier series,
\begin{equation}\label{eq:fourierSum}
P(y,\theta) \approx \sum_{m = -M}^M \sum_{n = -N}^N \widetilde{P}_{mn} e^{i(my + n\theta)},
\end{equation} 
where $2M+1$ is the maximum number of Fourier modes in the $y$ direction, $2N+1$ is the maximum number of modes in the $\theta$ direction, and the Fourier coefficients are defined by
\begin{equation}\label{eq:fourierCoeff}
\widetilde{P}_{mn} = \frac{1}{4\pi^2} \int_0^{2\pi} \int_0^{2\pi} P(y,\theta)e^{-i(my + n\theta)} {\rm d} y {\rm d} \theta.
\end{equation}
Taking the Fourier transform of Eq.~\eqref{eq:fp}, we obtain
\begin{align}
& \frac{v_0 m}{2} \left( \Pt_{m,n-1} - \Pt_{m,n+1} \right) + \frac{n}{4}  \bigg[ \Pt_{m-1,n} - \Pt_{m+1,n} \nonumber \\ \label{eq:fourierFP}
& - \frac{\alpha}{2} \left(\Pt_{m-1,n-2} + \Pt_{m-1,n+2} - \Pt_{m+1,n-2} - \Pt_{m+1,n+2} \right) \bigg] + \frac{\sigma^2 n^2}{2} \Pt_{mn} = 0.
\end{align}
Equation \eqref{eq:fourierFP} consists of $(2M+1)(2N+1)$ linear equations for the $\widetilde{P}_{mn}$, one for each $(m,n)$ of our truncated Fourier expansion.
For the combinations of $(m,n)$ such that Eq.~\eqref{eq:fourierFP} contains Fourier coefficients of higher modes than those retained by our truncated expansion \eqref{eq:fourierSum}, these higher order coefficients are simply dropped from Eq.~\eqref{eq:fourierFP}.
For $(m,n) = (0,0)$, Eq.~\eqref{eq:fourierFP} gives $0=0$.
We replace this equation by considering Eq.~\eqref{eq:fourierCoeff} with $(m,n) = (0,0)$, and applying the normalization condition $\int P {\rm d}y {\rm d} \theta = 1$.
This yields
\begin{equation}\label{eq:fourierNorm}
\widetilde{P}_{00} = \frac{1}{4 \pi^2}.
\end{equation}
Equation \eqref{eq:fourierNorm} combined with Eq.~\eqref{eq:fourierFP} for $(m,n) \neq (0,0)$ provides $(2M+1)(2N+1)$ linear equations for the $\widetilde{P}_{mn}$, which we formulate in matrix form and solve in Matlab \cite{code}.
For most of the calculations presented in the paper, we take $M=N=50$.
The one exception is the density plotted in Fig.~\ref{fig:Pyth}a, where we take $M=100$ in order to accurately resolve the small-scale variations in the $y$ direction.

\section*{Author Contributions}
SB conceived and designed the research.
SB, KF, and NB performed the analytical and numerical calculations.
All authors contributed to the interpretation of the results and data.
SB wrote the first draft of the manuscript.
All authors contributed to the revision of the manuscript. 

\section*{Funding}
This study was supported by the National Science Foundation under grants
CMMI-1825379 and DMR-1806355.

\section*{Data Availability Statement}
The original contributions presented in the study are included in the article.
The code is available at \cite{code}.
Further inquiries can be directed to the corresponding authors.

\bibliographystyle{frontiersinHLTH_FPHY}

\begin{thebibliography}{25}
\expandafter\ifx\csname natexlab\endcsname\relax\def\natexlab#1{#1}\fi
\expandafter\ifx\csname urlstyle\endcsname\relax
  \expandafter\ifx\csname doi\endcsname\relax
  \def\doi#1{doi:\discretionary{}{}{}#1}\fi \else
  \expandafter\ifx\csname doi\endcsname\relax
  \def\doi{doi:\discretionary{}{}{}\begingroup \urlstyle{rm}\Url}\fi \fi
\expandafter\ifx\csname selectlanguage\endcsname\relax
  \def\selectlanguage#1{}\fi

\bibitem[{Rusconi et~al.(2014)Rusconi, Guasto, and Stocker}]{Rusconi2014}
Rusconi R, Guasto JS, Stocker R.
\newblock {Bacterial transport suppressed by fluid shear}.
\newblock {\em Nat Phys\/} {\bf 10} (2014) 212--217.
\newblock \doi{10.1038/nphys2883}.

\bibitem[{Ebbens and Howse(2010)}]{Ebbens2010}
Ebbens SJ, Howse JR.
\newblock {In pursuit of propulsion at the nanoscale}.
\newblock {\em Soft Matter\/} {\bf 6} (2010) 726--738.
\newblock \doi{10.1039/b918598d}.

\bibitem[{Sanchez et~al.(2012)Sanchez, Chen, Decamp, Heymann, and
  Dogic}]{Sanchez2012}
Sanchez T, Chen DT, Decamp SJ, Heymann M, Dogic Z.
\newblock {Spontaneous motion in hierarchically assembled active matter}.
\newblock {\em Nature\/} {\bf 491} (2012) 431--434.
\newblock \doi{10.1038/nature11591}.

\bibitem[{Torney and Neufeld(2007)}]{Torney2007}
Torney C, Neufeld Z.
\newblock {Transport and aggregation of self-propelled particles in fluid
  flows}.
\newblock {\em Phys Rev Lett\/} {\bf 99} (2007) 078101.
\newblock \doi{10.1103/PhysRevLett.99.078101}.

\bibitem[{Khurana et~al.(2011)Khurana, Blawzdziewicz, and
  Ouellette}]{Khurana2011}
Khurana N, Blawzdziewicz J, Ouellette NT.
\newblock {Reduced transport of swimming particles in chaotic flow due to
  hydrodynamic trapping}.
\newblock {\em Phys Rev Lett\/} {\bf 106} (2011) 198104.
\newblock \doi{10.1103/PhysRevLett.106.198104}.

\bibitem[{Berman and Mitchell(2020)}]{Berman2020}
Berman SA, Mitchell KA.
\newblock {Trapping of swimmers in a vortex lattice}.
\newblock {\em Chaos\/} {\bf 30} (2020) 063121.
\newblock \doi{10.1063/5.0005542}.

\bibitem[{Ariel and Schiff(2020)}]{Ariel2020}
Ariel G, Schiff J.
\newblock {Conservative, dissipative and super-diffusive behavior of a particle
  propelled in a regular flow}.
\newblock {\em Phys D (Amsterdam, Neth)\/} {\bf 411} (2020) 132584.
\newblock \doi{10.1016/j.physd.2020.132584}.

\bibitem[{Berman et~al.(2021)Berman, Buggeln, Brantley, Mitchell, and
  Solomon}]{Berman2021a}
Berman SA, Buggeln J, Brantley DA, Mitchell KA, Solomon TH.
\newblock {Transport barriers to self-propelled particles in fluid flows}.
\newblock {\em Phys Rev Fluids\/} {\bf 6} (2021) L012501.
\newblock \doi{10.1103/PhysRevFluids.6.L012501}.

\bibitem[{Barry et~al.(2015)Barry, Rusconi, Guasto, and Stocker}]{Barry2015}
Barry MT, Rusconi R, Guasto JS, Stocker R.
\newblock {Shear-induced orientational dynamics and spatial heterogeneity in
  suspensions of motile phytoplankton}.
\newblock {\em J R Soc, Interface\/} {\bf 12} (2015)
  20150791.
\newblock \doi{10.1098/rsif.2015.0791}.

\bibitem[{Vennamneni et~al.(2020)Vennamneni, Nambiar, and
  Subramanian}]{Vennamneni2020}
Vennamneni L, Nambiar S, Subramanian G.
\newblock {Shear-induced migration of microswimmers in pressure-driven channel
  flow}.
\newblock {\em J Fluid Mech\/} {\bf 890} (2020) A15.
\newblock \doi{10.1017/jfm.2020.118}.

\bibitem[{Z{\"{o}}ttl and Stark(2012)}]{Zottl2012}
Z{\"{o}}ttl A, Stark H.
\newblock {Nonlinear dynamics of a microswimmer in Poiseuille flow}.
\newblock {\em Phys Rev Lett\/} {\bf 108} (2012) 218104.
\newblock \doi{10.1103/PhysRevLett.108.218104}.

\bibitem[{Z{\"{o}}ttl and Stark(2013)}]{Zottl2013}
Z{\"{o}}ttl A, Stark H.
\newblock {Periodic and quasiperiodic motion of an elongated microswimmer in
  Poiseuille flow}.
\newblock {\em Eur Phys J E\/} {\bf 36} (2013) 4.
\newblock \doi{10.1140/epje/i2013-13004-5}.

\bibitem[{Santamaria et~al.(2014)Santamaria, {De Lillo}, Cencini, and
  Boffetta}]{Santamaria2014}
Santamaria F, {De Lillo} F, Cencini M, Boffetta G.
\newblock {Gyrotactic trapping in laminar and turbulent Kolmogorov flow}.
\newblock {\em Phys Fluids\/} {\bf 26} (2014) 111901.
\newblock \doi{10.1063/1.4900956}.

\bibitem[{Chen and Thiffeault(2020)}]{Chen2020}
Chen H, Thiffeault JL.
\newblock {Shape matters: A Brownian microswimmer in a channel}.
\newblock {\em J Fluid Mech\/} {\bf 916} (2021) A15.
\newblock \doi{10.1017/jfm.2021.144}.

\bibitem[{Jeffery(1922)}]{Jeffery1922}
Jeffery GB.
\newblock {The Motion of Ellipsoidal Particles Immersed in a Viscous Fluid}.
\newblock {\em Proc R Soc A\/} {\bf 102} (1922) 161--179.
\newblock \doi{10.1098/rspa.1922.0078}.

\bibitem[{Arguedas-Leiva and Wilczek(2020)}]{Arguedas-Leiva2020}
Arguedas-Leiva JA, Wilczek M.
\newblock {Microswimmers in an axisymmetric vortex flow}.
\newblock {\em New J Phys\/} {\bf 22} (2020) 053051.
\newblock \doi{10.1088/1367-2630/ab776f}.

\bibitem[{Strogatz(2018)}]{Strogatz}
Strogatz SH.
\newblock {\em {Nonlinear dynamics and chaos: with applications to physics,
  biology, chemistry, and engineering}\/} (CRC Press) (2018).

\bibitem[{Thiffeault and Guo(2021)}]{Thiffeault2021}
Thiffeault JL, Guo J.
\newblock {Shake your hips: An anisotropic active Brownian particle with a
  fluctuating propulsion force}.  (2021) arXiv:2102.11758 [cond--mat.soft].

\bibitem[{Hyon et~al.(2012)Hyon, Marcos, Powers, Stocker, and Fu}]{Hyon2012}
Hyon Y, Marcos, Powers TR, Stocker R, Fu HC.
\newblock {The wiggling trajectories of bacteria}.
\newblock {\em J Fluid Mech\/} {\bf 705} (2012) 58--76.
\newblock \doi{10.1017/jfm.2012.217}.

\bibitem[{Solon et~al.(2015)Solon, Cates, and Tailleur}]{Solon2015}
Solon AP, Cates ME, Tailleur J.
\newblock {Active brownian particles and run-and-tumble particles: A
  comparative study}.
\newblock {\em Eur Phys J: Spec Top\/} {\bf 224} (2015)
  1231--1262.
\newblock \doi{10.1140/epjst/e2015-02457-0}.

\bibitem[{Berman and Mitchell(2021)}]{Berman2021b}
Berman SA, Mitchell KA.
\newblock {Swimmer dynamics in externally-driven fluid flows: The role of
  noise}  (2021) 	arXiv:2108.10488 [physics.flu-dyn].

\bibitem[22]{code}
Our code used to solve the Fokker Planck equation and compute the quantities associated with the averaged drift-diffusion model is available at https://github.com/saberman52/KolmogorovFlowSwimmerDynamics.

\bibitem[{Hassler(2016)}]{Hassler2016}
Hassler U.
\newblock {\em {Stochastic Processes and Calculus}\/} (Cham: Springer) (2016).

\bibitem[{Freidlin and Wentzell(2012)}]{Freidlin2012}
Freidlin MI, Wentzell AD.
\newblock {\em {Random Perturbations of Dynamical Systems}\/} (New York, NY:
  Springer) (2012).
\newblock \doi{10.1007/978-3-642-25847-3}.

\bibitem[{Locsei and Pedley(2009)}]{Locsei2009}
Locsei JT, Pedley TJ.
\newblock {Run and tumble chemotaxis in a shear flow: The effect of temporal
  comparisons, persistence, rotational diffusion, and cell shape}.
\newblock {\em Bull Math Biol\/} {\bf 71} (2009) 1089--1116.
\newblock \doi{10.1007/s11538-009-9395-9}.

\bibitem[{Bearon and Hazel(2015)}]{Bearon2015}
Bearon RN, Hazel AL.
\newblock {The trapping in high-shear regions of slender bacteria undergoing
  chemotaxis in a channel}.
\newblock {\em J Fluid Mech\/} {\bf 771} (2015) R3.
\newblock \doi{10.1017/jfm.2015.198}.

\end{thebibliography}

\end{document}